\documentstyle[12pt,epsfig]{article}
\makeatletter
\def\@cite#1#2{\hbox{$^{\mbox{\the\scriptfont0 #1}}$}}
\def\@biblabel#1{#1\hfill}
\makeatother
\begin{document}
\renewcommand{\thefootnote}{\fnsymbol{footnote}}
\begin{flushleft}
section: Theoretical and Physical Chemistry \\
running title: Flux Correlation Approach
\end{flushleft}
\begin{center}
{\LARGE Flux Correlation Approach to Thermal Reactions and
Recombination Rates}
\end{center}
\vspace{1.5cm}
\begin{center}
\begin{Large}
Koichi Saito\footnote{ksaito@tohoku-pharm.ac.jp} \\
Tohoku Pharmaceutical University, Sendai 981-8558, Japan
\end{Large}
\end{center}
\vspace{1.0cm}
\begin{abstract}
The rate constants for recombination and exchange processes are
studied in terms of two different flux correlation approaches:
one is the Yamamoto approach, which is based on the linear response
theory, and the other is the Miller one. Using those approaches
we consider two exactly solvable cases, i.e., the free particle and
the parabolic potential models. Since the rate constants for recombination
and exchange processes are calculated by Laplace transforms of the
flux correlation functions, the two approaches give different results.
In the present calculation, we find that the rate constant in the
Yamamoto approach is larger than that in the Miller approach by
about 40\%  at low temperature ($\sim 100$ K) and high pressure ($\sim
1$ GPa). The difference is considerable
in the region where quantum effects dominate.
\end{abstract}
%
\newpage
%
%

When one wants to study a chemical reaction at the most detailed
level, it is necessary to calculate the Schr{\"o}dinger equation for a
state-to-state differential scattering cross section, which is
a function of total energy $E$ and total angular momentum $J$.
Such quantum reactive scattering cross sections have
actually been studied for simple chemical reactions, where a
time-dependent scattering formalism based on the $S$-matrix Kohn
variational approach\cite{kohn} or a coupled channel method in
hyperspherical coodinates\cite{hyper} has usually been used.

However, in chemical applications, there are many cases where only 
the (microcanonical or canonical) rate constant for a reaction is
needed. If the full, state-to-state scattering calculation has been
carried out, the rate constant is, of course, given by an average of
the cross sections.  If it is, however, {\em only} the rate constant
that is desired, such a complete calculation for all state-to-state
information is {\em not} economical. Furthermore, because of the rapid
growth of the number of open vibration-rotation channels with
increasing thermally accessible collision energies, the calculation of
a rate constant via exact quantum state-to-state calculations would not
be feasible even for a simple reaction.

The traditional way of evaluating a rate constant
is the transition state theory (TST).\cite{tst} However, TST is a 
{\em classical} theory and it is approximate, because it does not 
involve the effect of {\em recrossing}
of the system over the transition state dividing a potential surface.
A number of improvements have been proposed to take the
recrossing effect into account.\cite{cross,miller}

In the early 60's, Yamamoto\cite{yama} first formulated an exact
expression for the rate constant as an application of the general
statistical mechanical theory of irreversible process, which was
established by Kubo et al.\cite{kubo} and Mori.\cite{mori}
Later (in the early 70's),
Miller et al.\cite{miller,miller2} separately developed a method for the
rate constant using a time integral of the flux-flux autocorrelation function,
which is exact in the limit that the dynamics is extended to $t \to \infty$.
Because the flux-flux correlation is calculated via time-dependent
quantum mechanics, the feasibility of this approach depends on how
to evaluate the Hamiltonian, flux and time evolution operators for
the system.  The flux-flux autocorrelation function
method has been applied to a variety of chemical
reactions.\cite{appl}  In particular, the reaction of $H + H_2$ has
been studied intensively.\cite{h2}

It has recently been shown how a quantum mechanical version of the
Lindemann mechanism for collisional recombination 
\begin{eqnarray}
A + B &\rightleftharpoons& AB^* , \label{ab1} \\
AB^* + M &\to& AB + M , \label{ab2}
\end{eqnarray}
can be handled by the flux-flux autocorrelation function for
the $A-B$ collision.\cite{appl,lind,lind2}  Here the process is
affected by the bath gas $M$.  Some applications of this new
theory are listed in Ref.14. 
It is furthermore possible to generalize the formalism to include
chemical reactions as well as recombination:
\begin{eqnarray}
A + BC &\rightleftharpoons& ABC^* \to AB + C , \label{abc1} \\
ABC^* + M &\to& ABC + M . \label{abc2}
\end{eqnarray}
Equations (\ref{abc1}) and (\ref{abc2}) simultaneously describe 
the recombination
process ($A + BC \to ABC$) and the exchange reaction ($A + BC \to AB +
C$).\cite{lind2} This method has been applied to the interesting
(combustion) reactions ($O + OH \rightleftharpoons H + O_2$) and the
recombination reactions ($O + OH + M \to HO_2 + M \gets H + O_2 + M$).
Those reactions are very important in atmospheric chemistry.\cite{atm} 

The purpose of this paper is to study the difference between the way
proposed by Yamamoto,\cite{yama} which is based on the linear response
theory or the so-called Kubo formula,\cite{kubo} and the flux-flux
autocorrelation function method 
proposed by Miller et al.\cite{miller,miller2}
The two approaches can provide the same result to the rate
constant for a simple chemical reaction, because it is given in terms
of the integral of the flux-flux correlation function with respect to
time.\cite{miller,miller2}  However, the shapes of the correlation
functions calculated by the two methods 
are quite different from each other. For the recombination and exchange
reactions (like eqs.(\ref{abc1}) and (\ref{abc2})), the rate constants
in the two approaches could be different because they are evaluated
by Laplace transforms of the flux-flux correlation functions.
It is expected that the difference will appear in the region where quantum
effects dominate. 

In this paper, we first review the correlation function method briefly 
and show the difference between Yamamoto's and Miller's approaches
explicitly.  The rate constants for recombination and exchange
processes are also discussed.  Next, we study two exactly
solvable cases, i.e., the free particle and the parabolic potential
models.  Finally, the summary and conclusion are given.

\section{Flux-flux correlation approach to rate constants}
\label{sec:2}

\begin{figure}[t]
\begin{center}
\epsfig{file=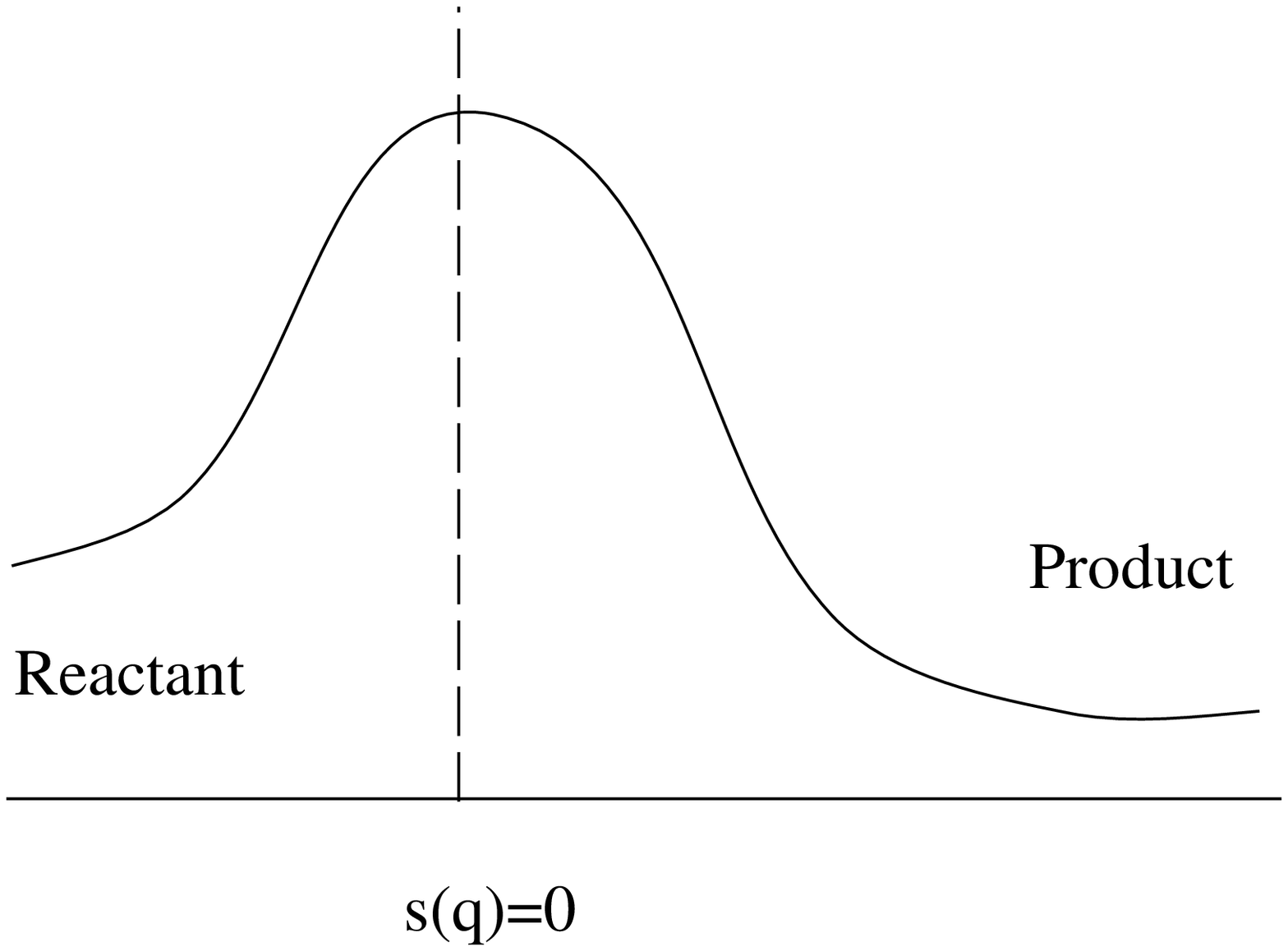,height=12cm}
\caption{K. Saito}
\label{f:reac}
\end{center}
\end{figure}

In the {\em classical} limit, a rate constant is generally given by 
an average of
the {\em flux} through some dividing surface that separates reactants
from products (see Fig.\ref{f:reac}).
The canonical rate constant is then given by\cite{appl}\footnote{
We use the natural unit, i.e., $h/2\pi = c = 1$.}
\begin{equation}
k_{cl}(T) = Q_r(T)^{-1}(2\pi)^{-f} \int d{\vec p}_1 \int d{\vec q}_1 
\ e^{-\beta H({\vec p}_1, {\vec q}_1)} F({\vec p}_1, {\vec q}_1)
{\cal P}({\vec p}_1, {\vec q}_1) ,  \label{cl}
\end{equation}
where $\beta^{-1} = k_BT$ ($T$, temperature) and
$({\vec p}_1, {\vec q}_1)$ provides the initial conditions of
the momenta and (reaction) coordinates for classical trajectories of the
system (consisting of $f$ degrees of freedom). The system is described by
the Hamiltonian $H({\vec p}_1, {\vec q}_1)$. Here $Q_r(T)$ is the
partition function per unit volume for the noninteracting reactants 
and $F$ is the flux factor 
which describes the trajectories crossing the dividing 
surface specified by $s({\vec q})=0$: 
\begin{equation}
F({\vec p}, {\vec q}) = \frac{d}{dt} h(s({\vec q})) = \delta(s({\vec q}))
v_s ,   \label{F}
\end{equation}
where $s({\vec q})$ is some function of position ${\vec q}$ that is
negative on the reactant side and positive on the product side.  Then, 
$h(s)$ is the step function, which is $+1 (0)$ for $s > (<) 0$,
and $v_s$ is the normal component of the velocity 
to the dividing surface $s({\vec q})$. 
The factor ${\cal P}$ in
eq.(\ref{cl}) involves all information of the dynamics and it is unity
when the trajectory is on the product side in the infinite future and
zero otherwise.  This implies that it is given by
\begin{equation}
{\cal P}({\vec p}_1, {\vec q}_1) = \lim_{t \to \infty} h(s({\vec q}(t)))
= \int_0^{\infty} dt \ \frac{d}{dt} h(s({\vec q}(t))) =
 \int_0^{\infty} dt \ F({\vec p}(t), {\vec q}(t)) ,   \label{P}
\end{equation}
where eq.(\ref{F}) is used. Thus, ${\cal P}$ provides the probability
that the trajectory lies on the product side of the dividing surface
at $t \to \infty$.  The rate constant then reads
\begin{equation}
Q_r k_{cl}(T) = \int_0^{\infty} dt \ C_{cl}(t) , \label{cl2}
\end{equation}
where
\begin{equation}
C_{cl}(t) = (2\pi)^{-f} \int d{\vec p}_1 \int d{\vec q}_1
\ e^{-\beta H({\vec p}_1, {\vec q}_1)} F({\vec p}_1, {\vec q}_1)
F({\vec p}(t), {\vec q}(t)) .  \label{cl3}
\end{equation}
This means that the rate constant is calculated by the time integral of
the flux-flux autocorrelation function $C_{cl}(t)$.

To take quantum effects into account, it is necessary to replace the
phase space integral by a quantum trace representation.
In the linear response theory,\cite{kubo} 
the response functon is usually defined as
\begin{equation}
\phi_{BA}(t) = -i {\rm tr}(\rho [A, B(t)]) , \label{res}
\end{equation}
where $\rho = e^{-\beta H}/{\rm tr}(e^{-\beta H})$ is the density operator
for an equilibrium state.  
After the perturbation by the operator $A$ at $t=0$, 
the response of the quantity $B(t)(= e^{itH}Be^{-itH})$ at
time $t$ in the system 
is described by the response function $\phi_{BA}(t)$.  For
the flux-flux autocorrelation function, one can identify
that $A = h(s)$ and $B(t) = F(t)$. Here $F(0)$ is the flux operator
at $t=0$, which is given by\cite{miller2}
\begin{equation}
F(0) = i [H, h(s)] = \frac{1}{2} \left[ \frac{p}{m}\delta(s) +
\delta(s)\frac{p}{m} \right] , \label{ff}
\end{equation}
with $p$ the momentum operator and $m$ the reduced mass of the system.
Thus, the response function for the rate constant is 
\begin{equation}
\phi_{Fh}(t) = -i {\rm tr}(\rho [h(s), F(t)]) . \label{resf}
\end{equation}
Using the Kubo identity\cite{kubo}
\begin{equation}
[A, e^{-\beta H}] = e^{-\beta H} \int_0^{\beta} d\lambda 
\ e^{\lambda H} [H,A] e^{-\lambda H} , \label{kuboi}
\end{equation}
the response function reads
\begin{equation}
\phi_{Fh}(t) 
= \int_0^{\beta} d\lambda \ {\rm tr}(\rho F(-i\lambda) F(t)) , \label{resf3}
\end{equation}
where eq.(\ref{ff}) is used and
$F(-i\lambda)=e^{\lambda H}F(0)e^{-\lambda H}$.

In the linear response theory, the relaxation function $\Phi_{BA}$ is
defined as
\begin{equation}
\Phi_{BA}(t) = \lim_{\epsilon \to 0} \int_t^\infty ds \ \phi_{BA}(s)
e^{-\epsilon s} . \label{relax}
\end{equation}
The relaxation function for the rate constant is thus given by
\begin{equation}
\Phi_{Fh}(t) = \int_t^\infty ds \ \phi_{Fh}(s) 
= \int_t^\infty ds \int_0^{\beta} d\lambda \ {\rm tr}
(\rho F(-i\lambda) F(s)) , \label{relaxf3}
\end{equation}
where we assumed that the response function decreases rapidly 
as $t \to \infty$. The rate constant in quantum mechanics is now given
in terms of the relaxation function at $t=0$ 
\begin{equation}
Q_r k(T) = \beta^{-1} \Phi_{Fh}(0)
= \beta^{-1} \int_0^\infty dt \int_0^{\beta} d\lambda \ {\rm tr}
(\rho F(-i\lambda) F(t)) \equiv \int_0^\infty dt \ C(t) , \label{kqm}
\end{equation}
where the flux-flux autocorrelation function in quantum mechanics 
$C(t)$ is defined by 
\begin{equation}
C(t) = \beta^{-1} \int_0^{\beta} d\lambda \ {\rm tr}
(e^{-\beta H} F(-i\lambda) F(t)) . \label{cqm}
\end{equation}
In the present notation, $C(t)$ in eq.(\ref{cqm}) corresponds to 
the flux correlation
proposed by Yamamoto.\cite{yama}  We should note that there exists an 
integral with respect to $\lambda$ which stems from the Kubo
identity and that 
it is dispensable in the {\em classical} limit $\beta \to 0$.
Because it is more convenient to use the commutation relation in
eq.(\ref{resf}) rather than eq.(\ref{cqm}) in
actual calculations, we re-define the Yamamoto's correlation function by 
\begin{equation}
C^Y(t) = \frac{1}{i\beta} {\rm tr}(e^{-\beta H} [h(s), F(t)]) , \label{cy}
\end{equation}
where the superscript $Y$ stands for ``Yamamoto''. 

By contrast, in Miller's approach\cite{miller,miller2} the
variable $\lambda$ in the flux is fixed to be $\beta/2$ 
and the $\lambda$ integral is 
performed.  Thus, from eq.(\ref{cqm}) Miller's correlation function is
given by 
\begin{equation}
C^M(t) = {\rm tr}(F(0)e^{it_c^* H}F(0)e^{-it_cH}) , \label{cm}
\end{equation}
where $t_c = t - i\beta/2$ and the superscript $M$ stands for
``Miller''.  This modification certainly makes actual
calculations simple, because the flux operators are involved
symmetrically in the correlation function.  
In fact, Yamamoto's correlation function 
$C^Y(t)$ is {\em not} identical to $C^M(t)$. However, their
integrals with respect to time are identical to each other, and hence
they can provide the same rate constant. Therefore, the way of Miller et
al.\cite{miller,miller2} certainly has some distinct advantages in
actual numerical calculations.

\section{Rate constants for recombination and exchange reactions}
\label{sec:3}

It is possible to generalize the flux correlation approach to treat  
recombination and exchange reactions.\cite{appl,lind,lind2} 
It may be intuitive and useful to begin with the classical
description of the process again.  Let us consider the reaction of 
$A + BC \to AB + C$ and $ABC$ (see Fig.\ref{f:diag}).  
The classical rate constants for the
exchange ($A + BC \to AB + C$) and recombination ($A + BC \to ABC$)
reactions are again given by eq.(\ref{cl}), i.e., averages of the
flux $F_r({\vec p}_1, {\vec q}_1)$ and the probability 
${\cal P}({\vec p}_1, {\vec q}_1)$ over the Boltzmann distribution.
Here $F_r$ is the flux at the reactant dividing surface $s_r$ (see 
Fig.\ref{f:diag}):
\begin{equation}
F_r = \frac{d}{dt} h(s_r) = \delta(s_r) v_r .   \label{Fr}
\end{equation}
Note that $h(s_r)$ is again the step function, which is $0(1)$ for
position ${\vec q}$ to the left (right) of the dividing surface $s_r$, 
and that $v_r$ is the normal component of the velocity to the
surface $s_r$.  Similarly we define the step function for the product
dividing surface $s_p$ by $h(s_p)$ (see Fig.\ref{f:diag}), that is,
$h(s_p) = 0(1)$ for position ${\vec q}$ to the left (right) of the
dividing surface $s_p$. 
The difference of those step functions, $h_c({\vec q}) = h(s_r({\vec q}))
- h(s_p({\vec q}))$, is unity for position ${\vec q}$ between the
two dividing surfaces (i.e., in the ``compound'' region) and zero outside.

Because the probability of the system experiencing a deactivating
($ABC^* + M \to ABC + M$) collision with the bath gas $M$ can be evaluated
by $1-e^{- \eta t}$ at time $t$ ($\eta$ describes the frequency of
deactivating collisions and it depends on pressure $P$ and $T$ of the bath
gas), the recombination probability is estimated as 
\begin{equation}
{\cal P}_{rec} = 1 - e^{- \eta \tau} ,   \label{pr}
\end{equation}
where $\tau$ is the time the trajectory (it is on $s_r$ at $t=0$) is in
the compound region.  Thus, using $h_c$ and an integration by parts, 
${\cal P}_{rec}$ reads\cite{lind2} 
\begin{equation}
{\cal P}_{rec} = \int_0^\infty dt \ h_c({\vec q}(t)) \frac{d}{dt}
(1 - e^{- \eta t}) = \int_0^\infty dt \ (e^{- \eta t} -1)(F_r(t) -
F_p(t)) ,   \label{pr2}
\end{equation}
where $F_i(t)= {\dot h}_i({\vec q}(t))$ ($i = r$ or $p$). 

{}For the exchange reaction, the probability is
given by $e^{- \eta \tau_p}$, where $\tau_p$ is the time the trajectory
exists through the surface $s_p$.  
The probability is eventually obtained as\cite{lind2}
\begin{equation}
{\cal P}_{exc} = \int_0^\infty dt \ (1-h_p({\vec q}(t)) \frac{d}{dt}
e^{- \eta t} = \int_0^\infty dt \ e^{- \eta t} F_p(t) .   \label{exc}
\end{equation}

Inserting those probability functions into
eq.(\ref{cl}), the rate constants for the recombination and exchange
reactions are given by
\begin{eqnarray}
Q_r k_{cl}^{rec}(T,P) &=& \int_0^\infty dt \ e^{- \eta t} (C_{rr}^{cl}(t) -
C_{rp}^{cl}(t)) , \label{kclrec}  \\
Q_r k_{cl}^{exc}(T,P) &=& \int_0^\infty dt \ e^{- \eta t} C_{rp}^{cl}(t) ,
\label{kclexc}
\end{eqnarray}
where
\begin{eqnarray}
C_{rr}^{cl}(t) &=& (2\pi)^{-f} \int d{\vec p}_1 \int d{\vec q}_1
\ e^{-\beta H({\vec p}_1, {\vec q}_1)} F_r({\vec p}_1, {\vec q}_1)
F_r({\vec p}(t), {\vec q}(t)) ,  \label{clcrr} \\
C_{rp}^{cl}(t) &=& (2\pi)^{-f} \int d{\vec p}_1 \int d{\vec q}_1
\ e^{-\beta H({\vec p}_1, {\vec q}_1)} F_r({\vec p}_1, {\vec q}_1)
F_p({\vec p}(t), {\vec q}(t)) .  \label{clcrp}
\end{eqnarray}
Here the relation 
\begin{equation}
\int_0^\infty dt \ C_{rr}^{cl}(t) = \int_0^\infty dt \ C_{rp}^{cl}(t) 
\label{crel}
\end{equation}
holds because in the limit $\eta \to 0$ the recombination rate
should vanish.

The transcription of the rate constants to quantum mechanics simply
involves replacing the classical correlation functions by their
quantum mechanical counterparts.  As in the classical case, 
the rate constants for recombination and exchange processes in quantum
mechanics are thus given by
\begin{eqnarray}
Q_r k_{rec}^{Y,M}(T,P) &=& \int_0^\infty dt \ e^{- \eta t}
(C_{rr}^{Y,M}(t) -
C_{rp}^{Y,M}(t)) , \label{krec}  \\
Q_r k_{exc}^{Y,M}(T,P) &=& \int_0^\infty dt \ e^{- \eta t} C_{rp}^{Y,M}(t) .
\label{kexc}
\end{eqnarray}
Then, the flux-flux autocorrelation functions are given by
\begin{eqnarray}
C_{rr}^Y(T) &=& \frac{1}{i\beta} {\rm tr}(e^{-\beta H} [h(s_r), F_r(t)]) ,
\label{cyrr} \\
C_{rp}^Y(T) &=& \frac{1}{i\beta} {\rm tr}(e^{-\beta H} [h(s_r), F_p(t)]) ,
\label{cyrp}
\end{eqnarray}
in Yamamoto's approach, while
\begin{eqnarray}
C_{rr}^M(T) &=& {\rm tr}(F_r(0)e^{it_c^* H}F_r(0)e^{-it_cH}) ,
\label{cmrr} \\
C_{rp}^M(T) &=& {\rm tr}(F_r(0)e^{it_c^* H}F_p(0)e^{-it_cH}) ,
\label{cmrp}
\end{eqnarray}
in Miller's approach.  Note that $F_i(0)$ ($i = r$ or $p$) is
the quantum mechanical flux, which is again given by eq.(\ref{ff})
with $\delta(s_i)$ and $h(s_i)$, instead of $\delta(s)$ and $h(s)$.

Because the rate constant is calculated by the Laplace transform of the
flux-flux autocorrelation function, it is clear that the two 
approaches give different results. It is expected that
they will coincide with each other in the classical limit, 
but the difference
becomes large in the region where the $\lambda$ integration in
eq.(\ref{cqm}) cannot be ignored. 

\section{Numerical calculations}
\label{sec:4}

In this section we calculate the (canonical) rate constants for
recombination and exchange reactions using the Feynman path integral
technique.\cite{path}
A huge calculation is usually required to obtain the exact
matrix elemens of propagators for a realistic system.
Furthermore, it is necessary to consider some approximations
and numerical techniques like Monte Carlo samplings\cite{numeric} to
perform it. 
Because the aim of this paper is to show how the rate constant in the
Miller approach is different from that in the Yamamoto case, it would 
be more intuitive and useful to consider a simple system rather than
a complicated case.  We here study two analytically solvable cases: 
i.e., the free particle and the parabolic potential models in
one-dimension, and leave more elaborate calculations for
nontrivial cases for a forthcoming paper.

\subsection{Free particle case}
\label{subsec:free}

We first study the free particle case (see Fig.\ref{f:free}). 
The propagator for the free particle in a coordinate representation can
be easily calculated by the path integral.\cite{path}  The matrix
element of the flux operator $F_i (i= r$ or $p)$ in coordinate space is
also found easily for the free particle system.  For details, see
Appendix~A.

The flux-flux autocorrelation function in the Miller approach is
eventually given by
\begin{equation}
C_{rp}^M(t) = \frac{1}{4\pi(t^2+\beta^2/4)^{3/2}}
\left[ \frac{\beta}{2} + \frac{2mt^2d^2}{t^2+\beta^2/4} \right]
\exp\left[ - \frac{m \beta d^2}{2(t^2+\beta^2/4)} \right] , \label{fcmrp}
\end{equation}
with $d$ the distance between $s_r$ and $s_p$
(see Fig.\ref{f:free}).  Note that the correlation depends on {\em only}
the distance $d$ and is independent of positions $s_r$ and $s_p$, as 
it should be.  From this expression the correlation function
$C_{rr}^M(t)$ is easily obtained as
\begin{equation}
C_{rr}^M(t) = \frac{\beta}{8\pi(t^2+\beta^2/4)^{3/2}} . \label{fcmrr}
\end{equation}
Those correlation functions are illustrated by the dotted curves in
Figs.\ref{f:frr} and \ref{f:frp}, in which we define $C_1 = md^2/2\beta$
and take $C_1 = 1.0$ to illustrate the correlation functions clearly. 
The rate constant for the reaction without recombination (i.e., in the
limit $\eta \to 0$) can be obtained by the
integral of eq.(\ref{fcmrr}) with respect to time (see
eq.(\ref{kqm})): 
\begin{equation}
Q_r k(T) = \int_0^\infty dt \ C_{rr}^M(t) =
\frac{1}{2\pi\beta}. \label{ktfree} 
\end{equation}

By contrast, in the Yamamoto approach the flux-flux autocorrelation
function is given by (see Appendix~A)
\begin{eqnarray}
C_{rp}^Y(t) &=& \frac{1}{2\pi\beta^2\sqrt{2t(t^2+\beta^2)}}
\exp\left[ - \frac{m \beta d^2}{2(t^2+\beta^2)} \right] \nonumber \\
&\times& \left[ (\sqrt{t^2+\beta^2} + t)^{3/2} \sin X +
(\sqrt{t^2+\beta^2} - t)^{3/2} \cos X \right] , \label{fcyrp}
\end{eqnarray}
where $X = m \beta^2 d^2 / 2t(t^2+\beta^2)$. This 
is not identical to eq.(\ref{fcmrp}). In particular, at short time it is
divergent like $\sim 1/\sqrt{t}$ although it is integrable.  Note
that it again depends on only the distance $d$.  If we set $s_r=s_p$ (or
$d=0$), we obtain\cite{appl} 
\begin{equation}
C_{rr}^Y(t) = \frac{(\sqrt{t^2+\beta^2} - t)^{3/2}}
{2\sqrt{2t}\pi\beta^2(t^2+\beta^2)^{1/2}} , \label{fcyrr}
\end{equation}
and, as expected, we can find that for the usual rate constant the
Yamamoto correlation function gives $Q_r k(T) =
1/2\pi\beta$, which is the same as that in the Miller approach 
(see eq.(\ref{ktfree})). 
Those correlation functions are shown by the solid curves in
Figs.\ref{f:frr} and \ref{f:frp}. 

It can be seen from Figs.\ref{f:frr} and \ref{f:frp} that the interference
effect in the correlation is taken into account correctly in the
Yamamoto approach (although the vibrating behavior is inconvenient for
numerical calculations).
Contrastingly, in the Miller correlation function the interference is
averaged and the shape is quite smooth.  Thus, it is very convenient for
actual computation. For the usual rate constant, the two approaches
certainly give the same result, as we have seen above.

Next, we calculate the rate constants for recombination and exchange
reactions. 
The rate constants are given by eqs.(\ref{krec}) and (\ref{kexc}).  If
we define the $(rr)$- and $(rp)$-rate constants by 
\begin{eqnarray}
Q_r k_{rr}^{Y,M} &=& \int_0^\infty dt \ e^{- \eta t} C_{rr}^{Y,M}(t) ,
\label{krr}  \\
Q_r k_{rp}^{Y,M} &=& \int_0^\infty dt \ e^{- \eta t} C_{rp}^{Y,M}(t) ,
\label{krp}
\end{eqnarray}
the rate constants are given as $k_{rec}^{Y,M} = k_{rr}^{Y,M} -
k_{rp}^{Y,M}$ and $k_{exc}^{Y,M} = k_{rp}^{Y,M}$.

Then, the Miller approach gives
\begin{eqnarray}
Q_r k_{rr}^{M} &=& \frac{1}{8\pi\beta}
\int_0^\infty dx \ \frac{e^{-\alpha x}}{(x^2+1/4)^{3/2}} ,
\label{fmkrr}  \\
Q_r k_{rp}^{M} &=& \frac{1}{8\pi\beta}
\int_0^\infty dx \ e^{-\alpha x} \left[ \frac{1}{(x^2+1/4)^{3/2}}
+ \frac{8C_1x^2}{(x^2+1/4)^{5/2}} \right]
\exp \left(- \frac{C_1}{x^2+1/4} \right) ,
\label{fmkrp}
\end{eqnarray}
where $x (=t/\beta)$ is a dimensionless variable and $\alpha = \beta \eta$,
while in the Yamamoto approach we find 
\begin{eqnarray}
Q_r k_{rr}^{Y} &=& \frac{1}{2\sqrt{2}\pi\beta}
\int_0^\infty dx \ e^{-\alpha x} \frac{(\sqrt{x^2+1}-x)^{3/2}}
{\sqrt{x(x^2+1)}} ,
\label{fykrr}  \\
Q_r k_{rp}^{Y} &=& \frac{1}{2\sqrt{2}\pi\beta}
\int_0^\infty dx
\ \frac{e^{-\alpha x}}{\sqrt{x(x^2+1)}}
\exp\left[ - \frac{C_1}{x^2+1} \right] \nonumber \\
&\times& \left[ (\sqrt{x^2+1} + x)^{3/2} \sin X'  +
(\sqrt{x^2+1} - x)^{3/2} \cos X' \right] , 
\label{fykrp}
\end{eqnarray}
with $X'=C_1/x(x^2+1)$. 

In order to convert the collision frequency $\eta$ to more familiar
variables, we approximate the collisional deactivation rate constant by
an expression given by the hard sphere collision
theory. Furthermore, if one uses the ideal gas
expansion, the frequency can be expressed by\cite{lind2}
\begin{equation}
\eta = k_{deact}[M] = P \sqrt{\frac{2000}{T}} \times 10^{-11} ,
\label{eta}
\end{equation}
with $\eta$ in $fs^{-1}$, $P$ in Pa and $T$ in K. Then, we find
\begin{equation}
\alpha = \beta \eta \simeq 3.24 \times \frac{P}{T^{3/2}} \times
10^{-6} .
\label{alpha}
\end{equation}
The factor $C_1$ is also converted as
\begin{equation}
C_1 = \frac{md^2}{2 \beta} = 0.0103 \times ATd^2 ,
\label{c1}
\end{equation}
with $A$ the reduced mass of the system in atomic mass units 
and $d$ in {\AA}.
In this paper we consider a system which has a small reduced mass
(like $H + CO \to HCO$ or $H + O_2 \to HO_2$) to illustrate the
difference between the two approaches clearly. In the following
calculations, we thus take $A=2$ and $d=2${\AA} and vary $T$ and $P$.

Now we are in a position to show our results for the free particle 
case. First we define ratios
\begin{eqnarray}
R_{rr}(T,P) = k_{rr}^{Y}/k_{rr}^{M} , \label{fRrr} \\
R_{rp}(T,P) = k_{rp}^{Y}/k_{rp}^{M} . \label{fRrp}
\end{eqnarray}
Figures~\ref{f:fRrr} and \ref{f:fRrp} illustrate the two ratios for the
free particle case. Here 
we choose $T = 100 \sim 400$ K and $P = 0.1 \sim 1$ GPa.
(To check the accuracy of the present numerical calculation, we have
also evaluated the rate constant for the free particle with $\eta =
0$ and compared the result with the exact value given by eq.(\ref{ktfree}).  
It is confirmed that the numerical calculation is sufficiently accurate.)
In $R_{rr}$, the ratio is enhanced at low $T$ and high $P$, where
quantum effects dominate, as we first expected in section~\ref{sec:3}.
The ratio reaches $1.38$ at $T = 100$ K and $P = 1$ GPa. Hence, the
difference between the Yamomoto and the Miller approaches becomes 
rather large in the region of low $T$ and high $P$.  
This tendency can be seen 
clearly in the contour plot of $R_{rr}$.  On the contrary, in
$R_{rp}$ the ratio is reduced in the region where the quantum effect is
strong.  It is about $0.53$ at $T = 100$ K and $P = 1$ GPa. The contour
plot shows the decreasing behavior of $R_{rp}$ at low $T$ and
high $P$.

Combining the $(rr)$- and $(rp)$-rate constants, one can calculate the
ratio of the recombination rate constants, $R_{rec} =
k_{rec}^Y/k_{rec}^M$. 
The ratio is presented in Fig.\ref{f:fRrec}. (Note that
the ratio for the exchange process is given by $R_{exc}=
k_{exc}^Y/k_{exc}^M = R_{rp}$.) The behavior of $R_{rec}$ seems
similar to $R_{rr}$ and the ratio again reaches $1.38$ at $T = 100$ K
and $P = 1$ GPa. From the contour plot we can see that there is a
small difference between $R_{rec}$ and $R_{rr}$.

\subsection{Parabolic potential case}
\label{subsec:para}

The second example is a reaction which occurs under a harmonic
oscillator potential.  We suppose that the potential has a 
frequency $\omega$, the minimum point at $x_0$ with its value $V_0$,
and $s_r$ and $s_p$ are located symmetrically with respect to the
minimum point (see Fig.\ref{f:hod}).
The propagator for a particle moving under the potential can be 
found by the path integral.\cite{path}
The flux-flux autocorrelation function in the Miller approach is
then calculated by (for details, see Appendix~B)
\begin{eqnarray}
C_{rp}^M(t) &=& \frac{\kappa^2e^{-\beta V_0}}{4\pi\beta^2}
\left[ \frac{\sinh(\kappa/2) \cos u}{(\sinh^2(\kappa/2)+\sin^2 u)^{3/2}}
\right.
+ \kappa C_1 \left. \frac{(\cosh(\kappa/2)+\cos u)^2 \sin^2 u}
{(\sinh^2(\kappa/2)+\sin^2 u)^{5/2}} \right] \nonumber \\
&\times&
\exp\left[ - \kappa C_1 \frac{\sinh(\kappa/2)(\cosh(\kappa/2)+\cos u)}
{\sinh^2(\kappa/2)+\sin^2 u} \right] , \label{hcmrp}
\end{eqnarray}
where $\kappa = \omega \beta$ and $u = \omega t$.  Note that the
correlation does not depend on the position of the minimum point explicitly.
Similarly the correlation $C_{rr}^M$ is obtained as
\begin{eqnarray}
C_{rr}^M(t) &=& \frac{\kappa^2e^{-\beta V_0}}{4\pi\beta^2}
\left[ \frac{\sinh(\kappa/2)\cos u}{(\sinh^2(\kappa/2)+\sin^2 u)^{3/2}}
\right.
- \kappa C_1 \left. \frac{(\cosh(\kappa/2)-\cos u)^2 \sin^2 u}
{(\sinh^2(\kappa/2)+\sin^2 u)^{5/2}} \right]  \nonumber \\
&\times&
\exp\left[ - \frac{\kappa C_1 \sinh(\kappa/2)(\cosh(\kappa/2)-\cos u)}
{\sinh^2(\kappa/2)+\sin^2 u} \right] . \label{hcmrr}
\end{eqnarray}
Those correlation functions are shown by the dotted curves in
Figs.\ref{f:horr} and \ref{f:horp} (we take $C_1=1.5$ and $\kappa = 1.0$
to illustrate the correlation functions clearly). In the limit
$\omega, V_0 
\to 0$ eqs.(\ref{hcmrp}) and (\ref{hcmrr}) are, of course, identical to
eqs.(\ref{fcmrp}) and (\ref{fcmrr}), respectively. 

After lengthy algebra, we can find the correlation functions in the
Yamamoto approach (for details, see Appendix~B).  For example, for $0 \le
u \le \pi$, the correlations are expressed by
\begin{eqnarray}
C_{rp}^Y(t) &=& \frac{\kappa e^{-\beta V_0}}{2\sqrt{2}\pi\beta^2
\sinh\kappa \sqrt{\sin u (\sinh^2\kappa + \sin^2 u)}}
\exp\left[ - \frac{\kappa C_1 \sinh\kappa (\cosh\kappa + \cos u)}
{2(\sinh^2\kappa + \sin^2 u)} \right] \nonumber \\
&\times& \left[ Z_- \cos\left(\frac{\kappa C_1}{2} Y_+ \right)
 + Z_+ \sin\left(\frac{\kappa C_1}{2} Y_+ \right) \right] , \label{hcyrp}
\end{eqnarray}
and
\begin{eqnarray}
C_{rr}^Y(t) &=& \frac{\kappa e^{-\beta V_0}}{2\sqrt{2}\pi\beta^2
\sinh\kappa \sqrt{\sin u (\sinh^2\kappa + \sin^2 u)}}
\exp\left[ - \frac{\kappa C_1 \sinh\kappa (\cosh\kappa - \cos u)}
{2(\sinh^2\kappa + \sin^2 u)} \right] \nonumber \\
&\times& \left[ Z_- \cos\left(\frac{\kappa C_1}{2} Y_- \right)
 - Z_+ \sin\left(\frac{\kappa C_1}{2} Y_- \right) \right] , \label{hcyrr}
\end{eqnarray}
where
\begin{eqnarray}
Z_{\pm} &=&
(\sqrt{\sinh^2\kappa + \sin^2 u} \pm \sin u)
(\sqrt{\sinh^2\kappa + \sin^2 u} \pm \cosh\kappa \sin u)^{1/2},
\label{zpm} \\
Y_\pm &=& \frac{(1 \pm \cos u) \sinh^2\kappa - (\cosh\kappa -1) \sin^2 u}
{\sin u (\sinh^2\kappa + \sin^2 u)} .  \label{xy}
\end{eqnarray}
Note that in the limit $\omega, V_0 \to 0$ the correlation functions 
approach those in the case of the free particle. 
Those correlation functions are also illustrated by the solid curves in
Figs.\ref{f:horr} and \ref{f:horp}.

One can see from the figures that the correct behavior of the correlation
function is quite complicated and it is {\em divergent} (like 
$\sim 1/\sqrt{u}$) at $u=0, \pi, 2\pi,
\cdots$. However, the Miller correlation function is smooth {\em
everywhere}
and it {\em never} diverges. All those functions are periodical
because of the harmonic oscillator potential and, as expected, the integral of
the correlation function over one period vanishes.

The rate constants for recombination and
exchange reactions are calculated by Laplace transforms of the correlation
functions.  In the Miller approach, 
the $(rr)$- and $(rp)$-rate constants are given by
\begin{eqnarray}
Q_r k_{rr}^M &=& \frac{\kappa e^{-\beta V_0}}{4\pi\beta} \int_0^\infty du
\left[ \frac{\sinh(\kappa/2)\cos u}{(\sinh^2(\kappa/2)+\sin^2 u)^{3/2}}
\right.
- \kappa C_1 \left. \frac{(\cosh(\kappa/2)-\cos u)^2 \sin^2 u}
{(\sinh^2(\kappa/2)+\sin^2 u)^{5/2}} \right]  \nonumber \\
&\times&
\exp\left[- \frac{\alpha}{\kappa}u
- \frac{\kappa C_1 \sinh(\kappa/2)(\cosh(\kappa/2)-\cos u)}
{\sinh^2(\kappa/2)+\sin^2 u} \right] . \label{hkmrr}
\end{eqnarray}
and
\begin{eqnarray}
Q_r k_{rp}^M &=& \frac{\kappa e^{-\beta V_0}}{4\pi\beta} \int_0^\infty du
\left[ \frac{\sinh(\kappa/2) \cos u}{(\sinh^2(\kappa/2)+\sin^2 u)^{3/2}}
\right.
+ \kappa C_1 \left. \frac{(\cosh(\kappa/2)+\cos u)^2 \sin^2 u}
{(\sinh^2(\kappa/2)+\sin^2 u)^{5/2}} \right] \nonumber \\
&\times&
\exp\left[-  \frac{\alpha}{\kappa}u
- \kappa C_1 \frac{\sinh(\kappa/2)(\cosh(\kappa/2)+\cos u)}
{\sinh^2(\kappa/2)+\sin^2 u} \right] . \label{hkmrp}
\end{eqnarray}
Similarly, we can obtain the Yamamoto rate constants. Because the
expression of the rate constant is, however, lengthy, we do not 
write it explicitly here. (See Appendix B.)

Now we show our results of the parabolic potential case.
In Figs.\ref{f:hRrr} and \ref{f:hRrp}, the ratios $R_{rr}$ and
$R_{rp}$ are illustrated. In the present calculation, we fix
$\kappa$ to be $0.05$, which means that the potential energy of the
harmonic oscillator is much weaker (about 5\%) than the typical thermal
energy $\beta^{-1}$.  We should note that the frequency $\omega$ is
varied so as to keep $\kappa = 0.05$ at each $T$. If we set 
$\kappa$ to be smaller than 0.05 (for example, $\kappa = 0.01$), the
ratio, as it should, becomes close to that of the free
particle case. 

In Fig.\ref{f:hRrr}, the ratio is again enhanced at
low $T$ and high $P$, which is similar to the result of the free particle
case. The ratio at $(T, P) = (100{\rm K}, 1{\rm GPa})$ is about 1.38.
One distinct feature in the parabolic case is an enhancement of the ratio
in the region of high $T$ and low $P$. This can also be seen in the
contour plot.  In such a region, the power $\alpha$ 
appearing in Laplace transform for the rate constant is small, and hence
the rate constant at high $T$ and low $P$ is more influenced by the correlation
function at large $t$ than that at other $T$ and $P$, that is, the
rate constant is considerably affected by the (second) complicated 
structure around $u \sim \pi$ in the correlation function (see
Figs.\ref{f:horr} and \ref{f:horp}). This is
the reason why the enhancement at high $T$ and low $P$ appears in the
ratio. The ratio at $(T, P) = (400{\rm K}, 0.1{\rm GPa})$ is about 1.11.
In Fig.\ref{f:hRrp} a similar tendency can be seen: the ratio
is reduced at low $T$ and high $P$ ($R_{rp}=0.58$ at $(T, P) = (100{\rm K},
1{\rm GPa})$), which is similar to the free particle case, while it is
enhanced at high $T$ and low $P$ ($R_{rp}=1.43$ at $(T, P) = (400{\rm K},
0.1{\rm GPa})$).

Combining the $(rr)$- and $(rp)$-rate constants, we can calculate the
rate constant for the recombination process; this is presented in
Fig.\ref{f:hRrec}. In $R_{rec}$, the ratio is enhanced at low $T$ and
high $P$ ($R_{rec}=1.38$ at $(T, P) = (100{\rm K}, 1{\rm GPa})$),
while it is reduced at high $T$ and low $P$ ($R_{rec}=0.38$ at $(T, P) =
(400{\rm K}, 0.1{\rm GPa})$).  This behavior is also seen in the
contour plot. 

\section{Summary and Conclusion}
\label{sec:5}

The exact quantum mechanical expression for thermal reaction rates can be
formulated by the linear response theory,\cite{kubo,mori} which
Yamamoto first discussed in the early 60's.\cite{yama} 
Later, in the early 70's, Miller et al.\cite{miller,miller2}
have independently proposed a more convenient way to perform numerical
computation, which can provide the exact rate constant in the limit that
the dynamics of the system is extended to $t \to \infty$.

We have studied the difference between the two approaches in thermal
reactions which involve exchange and recombination processes. Because
the rate constants in those reactions are calculated by 
Laplace transforms of the flux-flux autocorrelation functions, the
results evaluated by the two approaches are different. In this paper,
we have considered two solvable cases, i.e., the free
particle and parabolic potential models, to demonstrate the difference
intuitively.  We have found that the shapes of the correlation
functions are quite different in the two approaches and that the
difference of the rate constants appears in the region where quantum
effects dominate.  In both the free and parabolic cases, the rate
constant for recombination in the Yamamoto approach is larger than
that in the Miller approach; the enhancement becomes about 40\%
at low temperature and high pressure. 

In conclusion, the Miller method is certainly an economical and 
powerful tool to
perform numerical calculations for thermal rates of realistic reactions.
However, it
may underestimate the rate constants for recombination and exchange
processes in the region where quantum effects dominate, because of neglecting
the $\lambda$ integral appearing in the Kubo identity.

\newpage
\underline{Appendix A --- free particle case}
\vspace{0.5cm}

The matrix element of the flux operator $F_i (i=r,$ or $p)$ in the
coordinate representation is given by\cite{miller2}
\begin{equation}
\langle u | F_i | u' \rangle = \frac{1}{2im}[\delta'(u-s_i)\delta(u'-s_i)
- \delta(u-s_i)\delta'(u'-s_i)] .  \label{flux}
\end{equation}
Using this expression, one can evaluate the flux-flux autocorrelation 
function in the Miller approach as 
\begin{eqnarray}
C_{rp}^M(T) &=& {\rm tr}(F_r(0)e^{it_c^* H}F_p(0)e^{-it_cH}) , \nonumber \\
&=& -\frac{1}{2m^2} \Re \left[ \frac{\partial}{\partial u} \langle u |
e^{-iHt_c} | u' \rangle^\star \frac{\partial}{\partial u'} \langle u |
e^{-iHt_c} | u' \rangle \right. \nonumber \\
&-& \left. \langle u | e^{-iHt_c} | u' \rangle^\star
\frac{\partial^2}{\partial u \partial u'}
\langle u | e^{-iHt_c} | u' \rangle \right]_{u=s_r, u'=s_p} ,
\label{acmrp}
\end{eqnarray}
where $t_c = t - i\beta/2$ and $\Re$ stands for taking the real part. 
On the other hand, the Yamamoto correlation function is given by 
\begin{eqnarray}
C_{rp}^Y(T) &=& \frac{1}{i\beta} {\rm tr}(e^{-\beta H} [h(s_r), F_p(t)]) ,
\nonumber \\
&=& \frac{1}{m\beta} \Im \int_{s_r}^\infty du
\left[ i \langle u | e^{-iHt_\beta} | u' \rangle^\star
\frac{\partial}{\partial u'}
\langle u | e^{-iHt} | u' \rangle \right. \nonumber \\
&-& \left. i \langle u | e^{-iHt} | u' \rangle
\frac{\partial}{\partial u'} \langle u | e^{-iHt_\beta} | u' \rangle^\star
\right]_{u'=s_p}, 
\label{acyrp}
\end{eqnarray}
where $t_\beta = t - i\beta$ and $\Im$ stands for the imaginary part.

With use of the path integral technique,\cite{path} 
the propagator for the free
particle at finite $\beta$ is calculated by
\begin{equation}
\langle u | e^{-iHt_\beta} | u' \rangle = \sqrt{\frac{m}{2\pi i}}
\frac{e^{i\theta/2}}{(t^2+\beta^2)^{1/4}} \exp \left[
\frac{m(it-\beta)}{2(t^2+\beta^2)}(u-u')^2 \right] , \label{afree}
\end{equation}
where
\begin{equation}
\cos \theta = \frac{t}{\sqrt{t^2+\beta^2}} \ \ \mbox{and} \ \
\sin \theta = \frac{\beta}{\sqrt{t^2+\beta^2}} . \label{theta}
\end{equation}

We can easily calculate the correlation functions 
using those expressions. The final results are explicitly presented in 
eqs.(\ref{fcmrp})--(\ref{fcyrr}).

\newpage
\underline{Appendix B --- parabolic potential case}
\vspace{0.5cm}

The propagator for a particle moving under the
harmonic oscillator potential (at finite $\beta$) which has the
minimum point at $(x_0, V_0)$ (see Fig.\ref{f:hod}) is evaluated by\cite{path}
\begin{eqnarray}
\langle u | e^{-iHt_\beta} | u' \rangle &=& \sqrt{
\left( \frac{m\omega}{2\pi i}\right)
\frac{\cosh\kappa \sin u + i \sinh\kappa \cos u}{\sinh^2\kappa
+ \sin^2 u}} e^{-\beta V_0} \nonumber \\
&\times& \exp \left[ i m \omega \left( \frac{\sin u \cos u
+ i \sinh\kappa \cosh\kappa}{2(\sinh^2\kappa + \sin^2 u)} \right) (u^2 + u'^2)
\right. \nonumber \\
&-& \left. i m \omega \left( \frac{\cosh\kappa \sin u + i \sinh\kappa \cos u}
{\sinh^2\kappa + \sin^2 u}\right) u u' - i V_0 t \right] ,
\label{ahoprp}
\end{eqnarray}
where $\kappa = \omega\beta$, $u = \omega t$ and $t_\beta = t - i\beta$.

Because it is easy to calculate the Miller correlation function using
eqs.(\ref{acmrp}) and (\ref{ahoprp}), we do not show its derivation here.
The final result is given in eqs.(\ref{hcmrp}) and (\ref{hcmrr}).
Instead, we explicitly present the Yamamoto correlation function.
It is a periodical function and the 
period is $4\pi$ (see Figs.\ref{f:horr} and \ref{f:horp}).  
Then, we divide it into 4 parts.  
Using eqs.(\ref{acyrp}) and (\ref{ahoprp}), we find: \\ 
(1) for $0 \le u \le \pi$, the results are given by eqs.(\ref{hcyrp})
and (\ref{hcyrr}). \\
(2) for $\pi < u \le 2\pi$,
\begin{eqnarray}
C_{rr}^Y(t) &=& \frac{\kappa e^{-\beta V_0}}{2\sqrt{2}\pi\beta^2
\sinh\kappa \sqrt{|\sin u| (\sinh^2\kappa + \sin^2 u)}}
\exp\left[ - \frac{\kappa C_1 \sinh\kappa (\cosh\kappa - \cos u)}
{2(\sinh^2\kappa+\sin^2 u)} \right] \nonumber \\
&\times& \left[ Z_+ \cos\left(\frac{\kappa C_1}{2} Y_- \right)
 - Z_- \sin\left(\frac{\kappa C_1}{2} Y_- \right) \right] , \label{hcyrr2}
\end{eqnarray}
and
\begin{eqnarray}
C_{rp}^Y(t) &=& \frac{\kappa e^{-\beta V_0}}{2\sqrt{2}\pi\beta^2
\sinh\kappa \sqrt{|\sin u| (\sinh^2\kappa + \sin^2 u)}}
\exp\left[ - \frac{\kappa C_1 \sinh\kappa (\cosh\kappa + \cos u)}
{2(\sinh^2\kappa+\sin^2 u)} \right] \nonumber \\
&\times& \left[ Z_+ \cos\left(\frac{\kappa C_1}{2} Y_+ \right)
 + Z_- \sin\left(\frac{\kappa C_1}{2} Y_+ \right) \right] , \label{hcyrp2}
\end{eqnarray}
where $Z_{\pm}$ and $Y_{\pm}$ are defined by eqs.(\ref{zpm})
and (\ref{xy}). \\
(3) for $2\pi < u \le 3\pi$,
\begin{eqnarray}
C_{rr}^Y(t) &=& \frac{\kappa e^{-\beta V_0}}{2\sqrt{2}\pi\beta^2
\sinh\kappa \sqrt{\sin u (\sinh^2\kappa + \sin^2 u)}}
\exp\left[ - \frac{\kappa C_1 \sinh\kappa (\cosh\kappa - \cos u)}
{2(\sinh^2\kappa+\sin^2 u)} \right] \nonumber \\
&\times& \left[ - Z_- \cos\left(\frac{\kappa C_1}{2} Y_- \right)
 + Z_+ \sin\left(\frac{\kappa C_1}{2} Y_- \right) \right] , \label{hcyrr3}
\end{eqnarray}
and
\begin{eqnarray}
C_{rp}^Y(t) &=& \frac{\kappa e^{-\beta V_0}}{2\sqrt{2}\pi\beta^2
\sinh\kappa \sqrt{\sin u (\sinh^2\kappa + \sin^2 u)}}
\exp\left[ - \frac{\kappa C_1 \sinh\kappa (\cosh\kappa + \cos u)}
{2(\sinh^2\kappa+\sin^2 u)} \right] \nonumber \\
&\times& \left[ - Z_- \cos\left(\frac{\kappa C_1}{2} Y_+ \right)
- Z_+ \sin\left(\frac{\kappa C_1}{2} Y_+ \right) \right] . \label{hcyr3}
\end{eqnarray}
(4) for $3\pi < u \le 4\pi$,
\begin{eqnarray}
C_{rr}^Y(t) &=& \frac{\kappa e^{-\beta V_0}}{2\sqrt{2}\pi\beta^2
\sinh\kappa \sqrt{|\sin u| (\sinh^2\kappa + \sin^2 u)}}
\exp\left[ - \frac{\kappa C_1 \sinh\kappa (\cosh\kappa - \cos u)}
{2(\sinh^2\kappa+\sin^2 u)} \right] \nonumber \\
&\times& \left[ - Z_+ \cos\left(\frac{\kappa C_1}{2} Y_- \right)
 + Z_- \sin\left(\frac{\kappa C_1}{2} Y_- \right) \right] , \label{hcyrr4}
\end{eqnarray}
and
\begin{eqnarray}
C_{rp}^Y(t) &=& \frac{\kappa e^{-\beta V_0}}{2\sqrt{2}\pi\beta^2
\sinh\kappa \sqrt{|\sin u| (\sinh^2\kappa + \sin^2 u)}}
\exp\left[ - \frac{\kappa C_1 \sinh\kappa (\cosh\kappa + \cos u)}
{2(\sinh^2\kappa+\sin^2 u)} \right] \nonumber \\
&\times& \left[ - Z_+ \cos\left(\frac{\kappa C_1}{2} Y_+ \right)
- Z_- \sin\left(\frac{\kappa C_1}{2} Y_+ \right) \right] . \label{hcyr4}
\end{eqnarray}

As shown in Figs.\ref{f:horr} and \ref{f:horp}, there is 
a symmetry property:
$C_{ri}^Y(0 \le u \le \pi) = - C_{ri}^Y(2\pi < u \le 3\pi)$ and
$C_{ri}^Y(\pi \le u \le 2\pi) = - C_{ri}^Y(3\pi < u \le 4\pi)$
($i = r$ or $p$). Thus, it is enough to calculate the correlation
function in the region of $0 \le u \le 2\pi$.  The rate constants are
calculated by Laplace transforms of those periodical
correlation functions.

\newpage
%

%

\newpage
\begin{center}
\underline{\large Figure captions}
\end{center}
\vspace{0.5cm}

\noindent Fig. 1: Sketch of a potential surface in one-dimensional
reaction versus the reaction coordinate $q$. 

\noindent Fig. 2: One-dimensional schematic diagram of the interaction
potential for the $A+BC \to AB+C$ reaction.
The compound region, $ABC$, is bounded by the dividing surfaces on
reactant ($s_r$) and product ($s_p$) sides. 

\begin{figure}[t]
\begin{center}
\epsfig{file=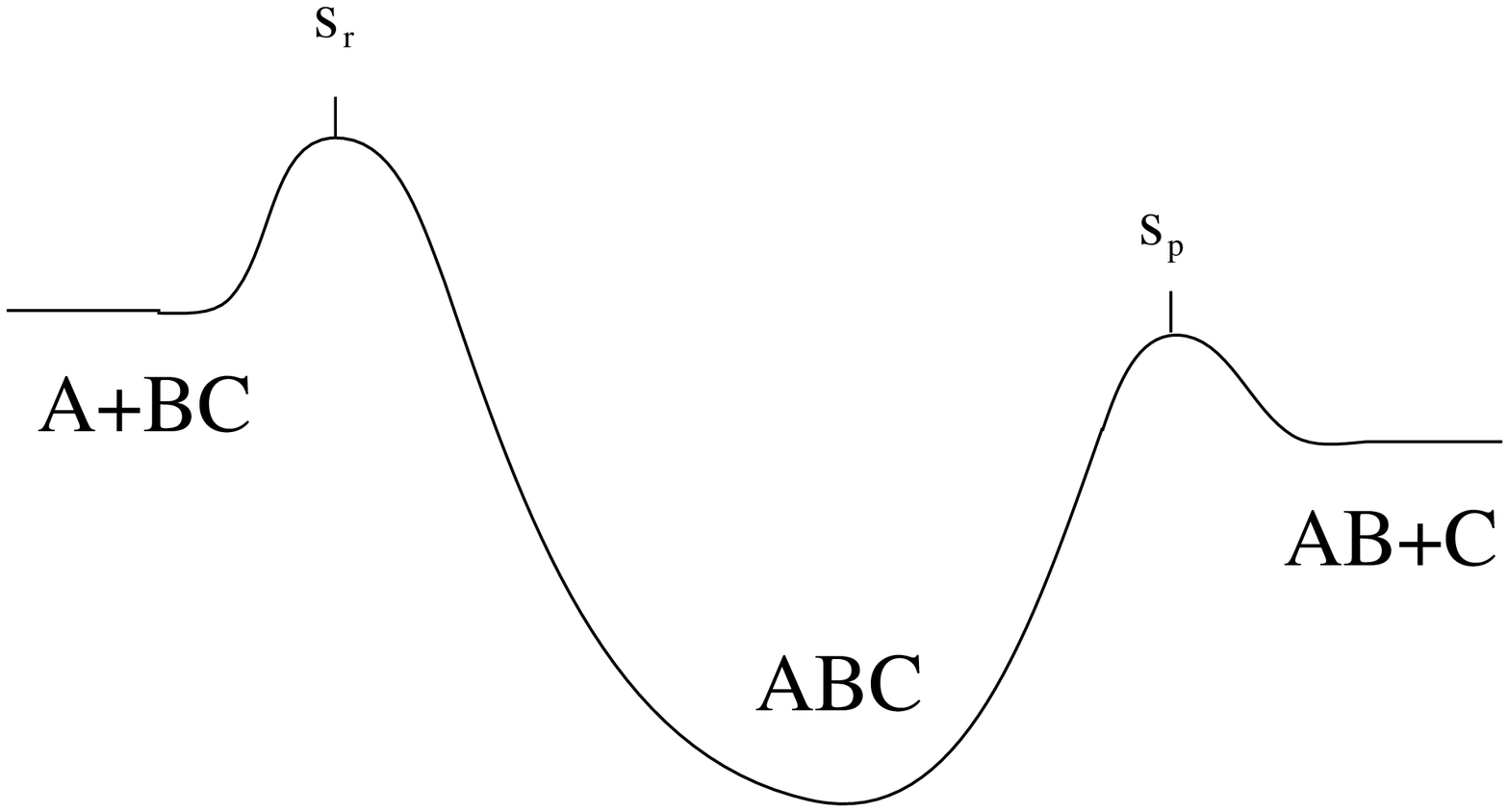,height=12cm}
\caption{K. Saito}
\label{f:diag}
\end{center}
\end{figure}

\begin{figure}[t]
\begin{center}
\epsfig{file=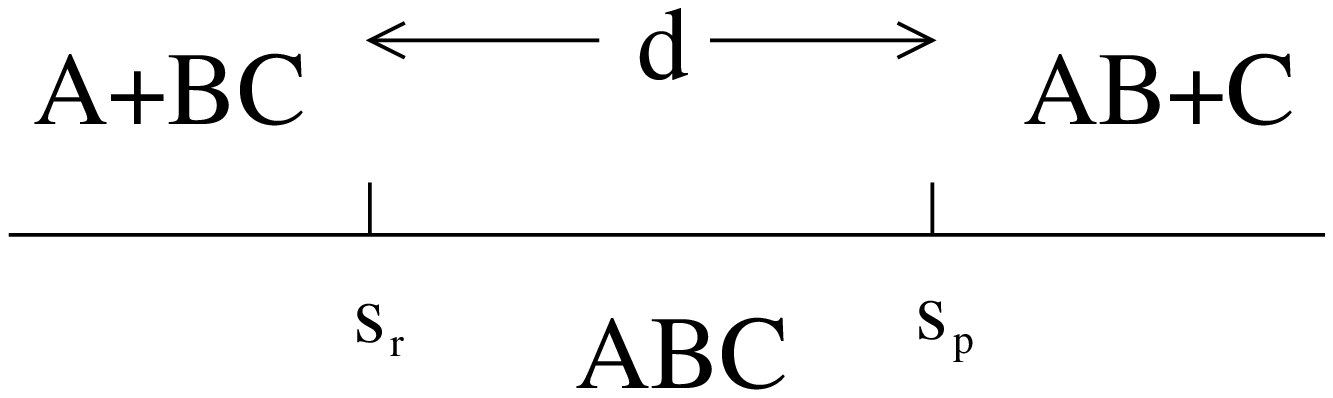,height=12cm}
\caption{K. Saito}
\label{f:free}
\end{center}
\end{figure}

\begin{figure}[t]
\begin{center}
\epsfig{file=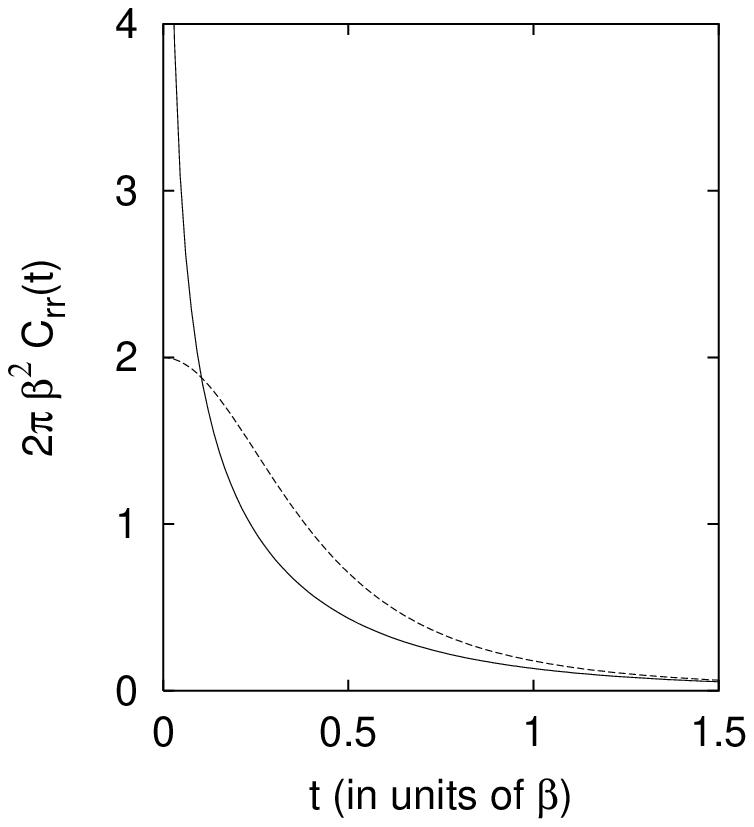,height=11cm}
\caption{K. Saito}
\label{f:frr}
\end{center}
\end{figure}

\begin{figure}[t]
\begin{center}
\epsfig{file=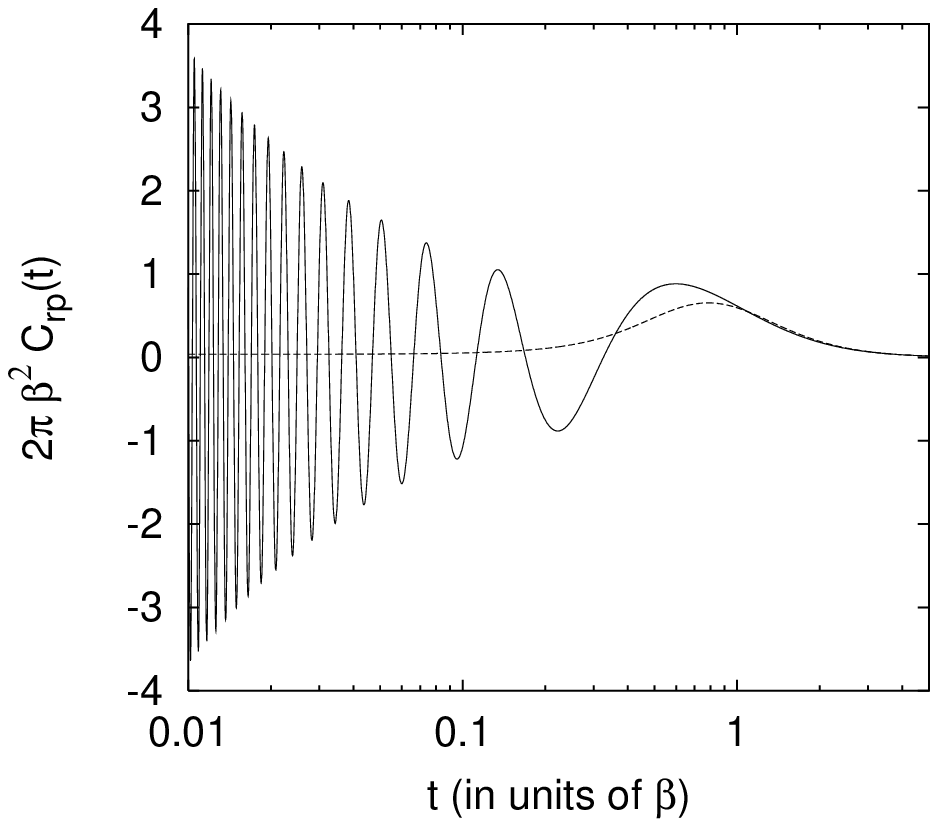,height=11cm}
\caption{K. Saito}
\label{f:frp}
\end{center}
\end{figure}

\noindent Fig. 3: Same as Fig.\ref{f:diag} but for the free
particle case. 

\noindent Fig. 4: Correlation function $C_{rr}(t)$ for the free particle.
The dotted curve is for $C_{rr}^M(t)$, while the solid curve is
for $C_{rr}^Y(t)$. We take $C_1 = md^2/2\beta = 1.0$. 

\noindent Fig. 5: Correlation function $C_{rp}(t)$ for the free particle.
The dotted curve is for $C_{rp}^M(t)$, while the solid curve is
for $C_{rp}^Y(t)$.  We take $C_1 = md^2/2\beta = 1.0$. 

\noindent Fig. 6: Ratio of the $(rr)$-rate constants in the free particle case
(top) and the contour plot (bottom).  In the contour plot, the top
dotted curve corresponds to $R_{rr}=1.05$ and the other curves are plotted
at intervals of 0.05. The bottom and right dotted curve is thus for
$R_{rr}=1.35$. 

\begin{figure}[t]
\begin{center}
\epsfig{file=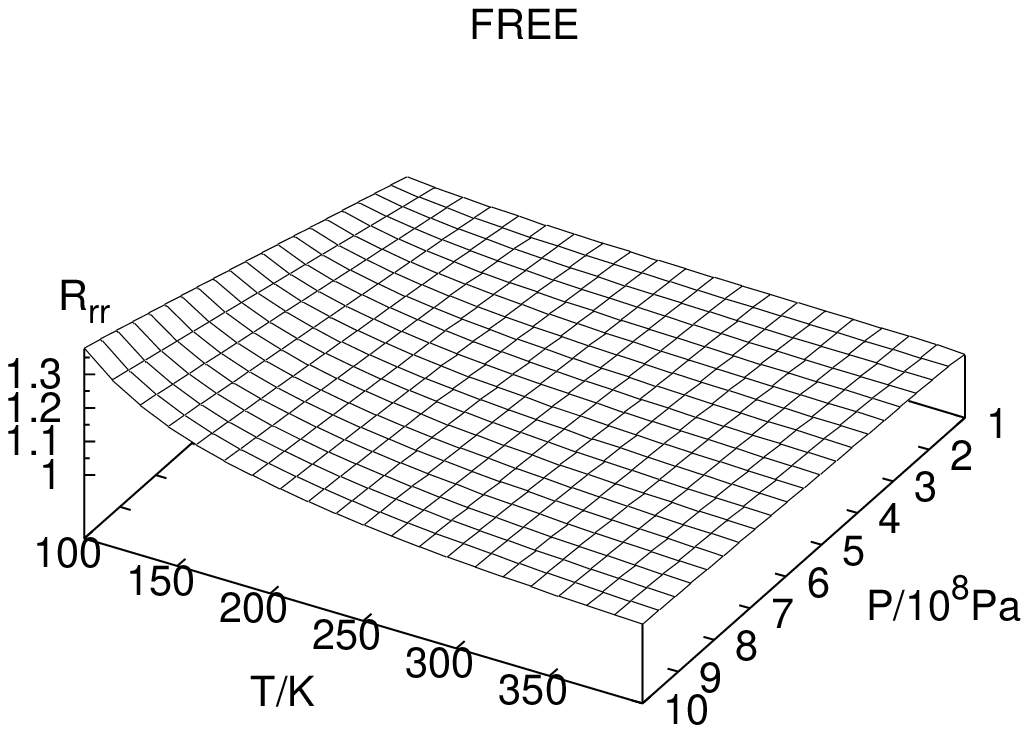,height=10cm}
\epsfig{file=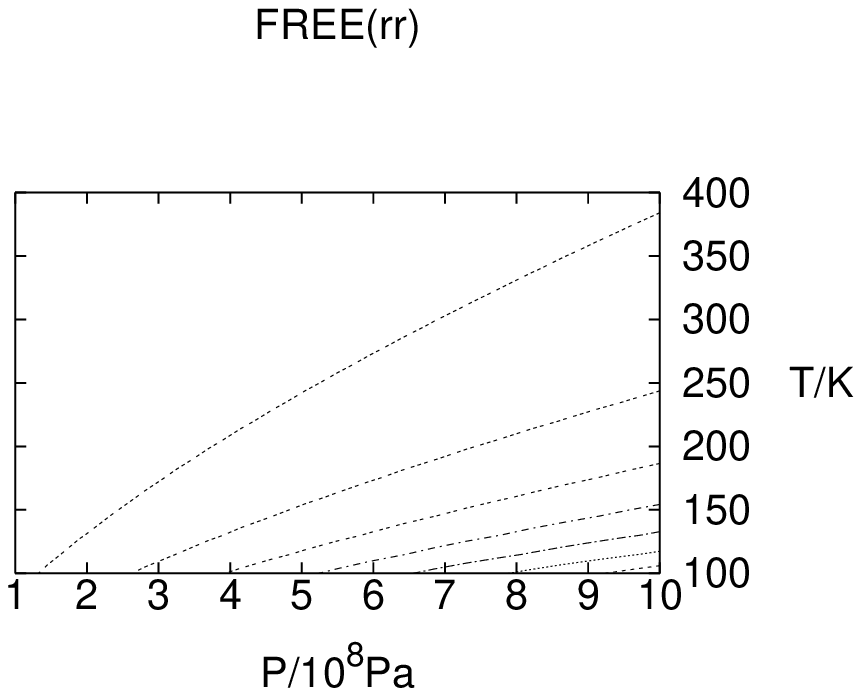,height=10cm}
\caption{K. Saito}
\label{f:fRrr}
\end{center}
\end{figure}

\begin{figure}[t]
\begin{center}
\epsfig{file=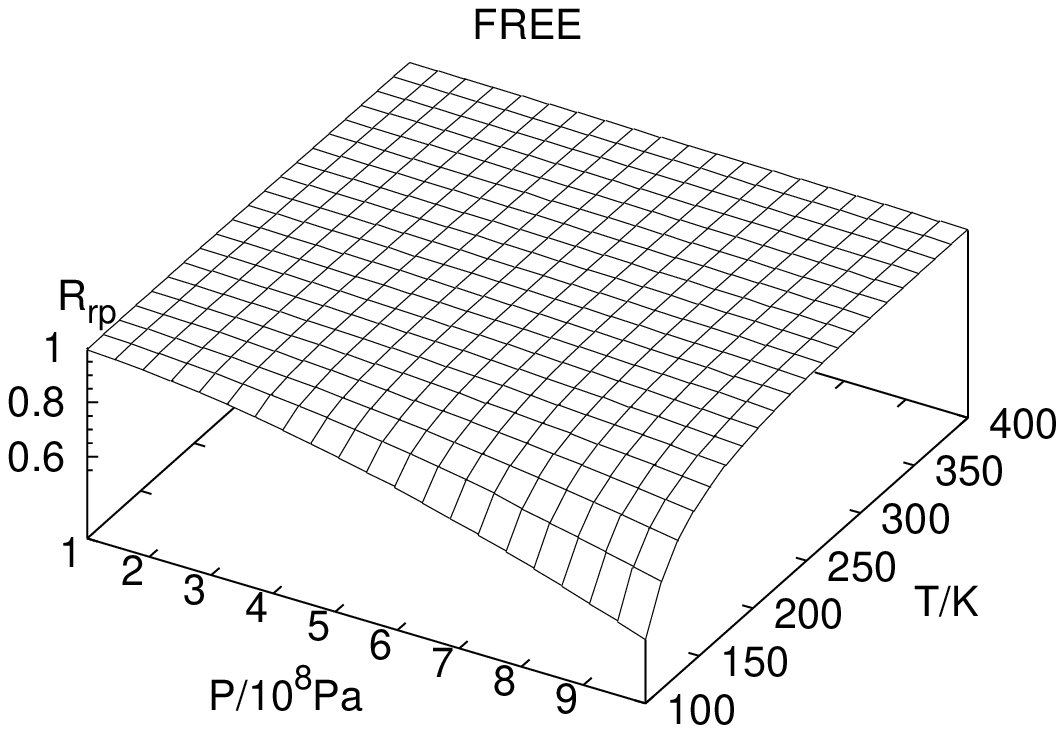,height=10cm}
\epsfig{file=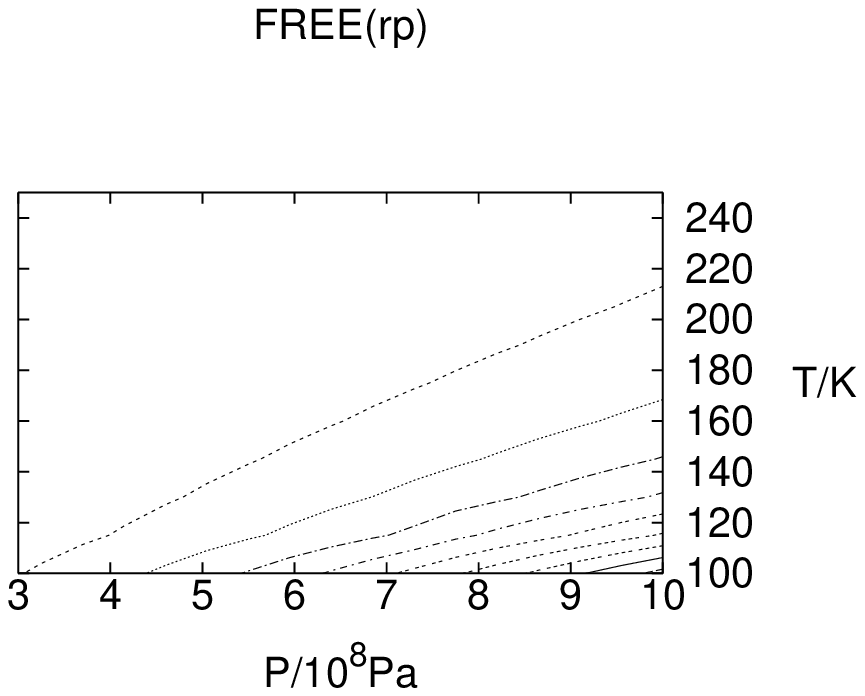,height=10cm}
\caption{K. Saito}
\label{f:fRrp}
\end{center}
\end{figure}

\begin{figure}[t]
\begin{center}
\epsfig{file=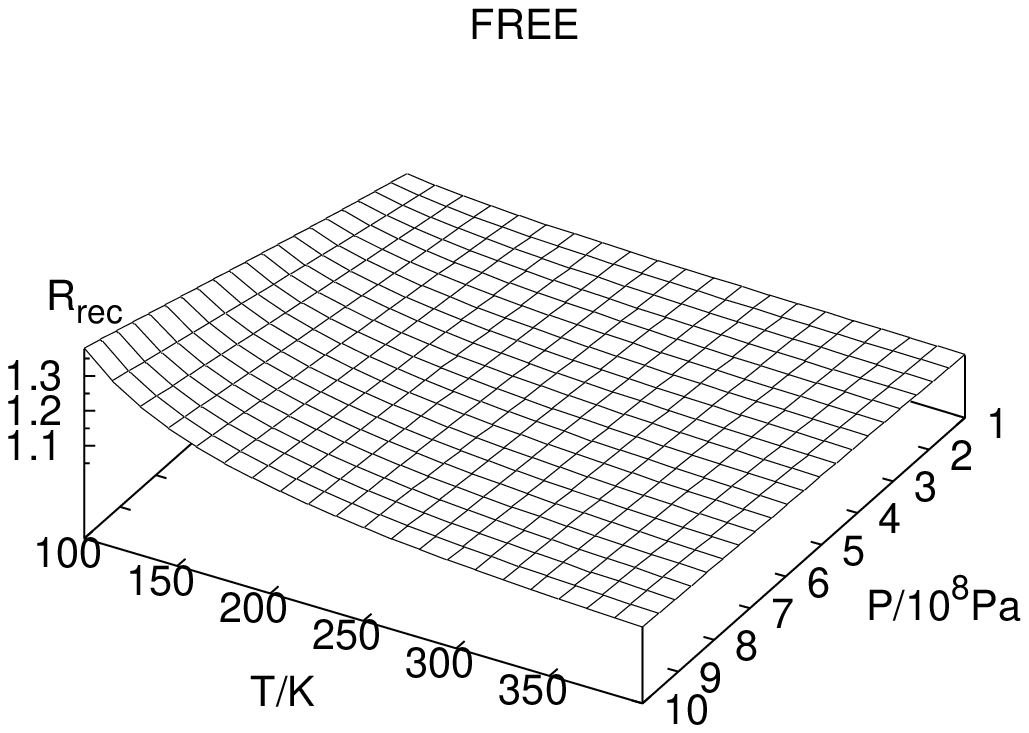,height=10cm}
\epsfig{file=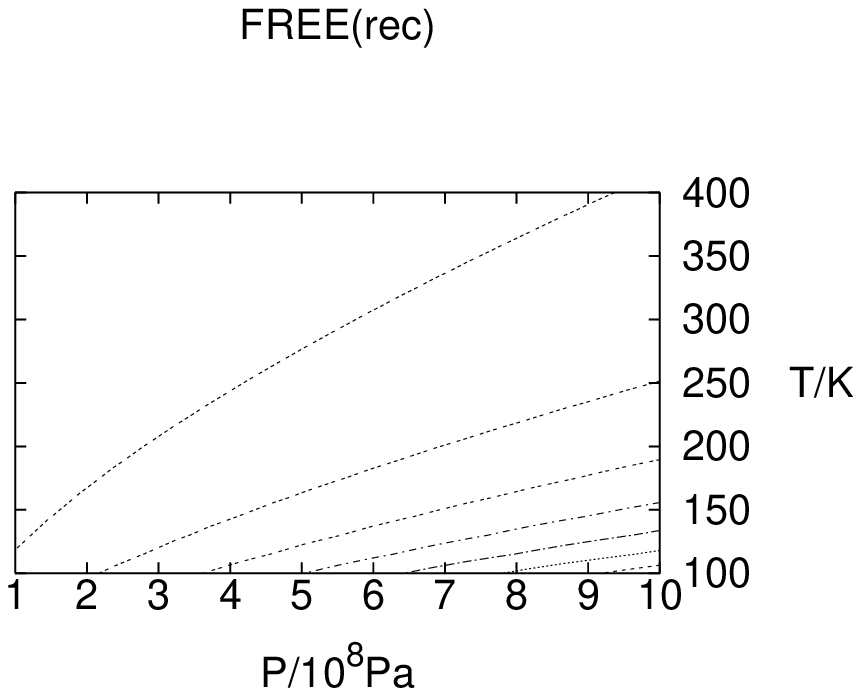,height=10cm}
\caption{K. Saito}
\label{f:fRrec}
\end{center}
\end{figure}

\begin{figure}[t]
\begin{center}
\epsfig{file=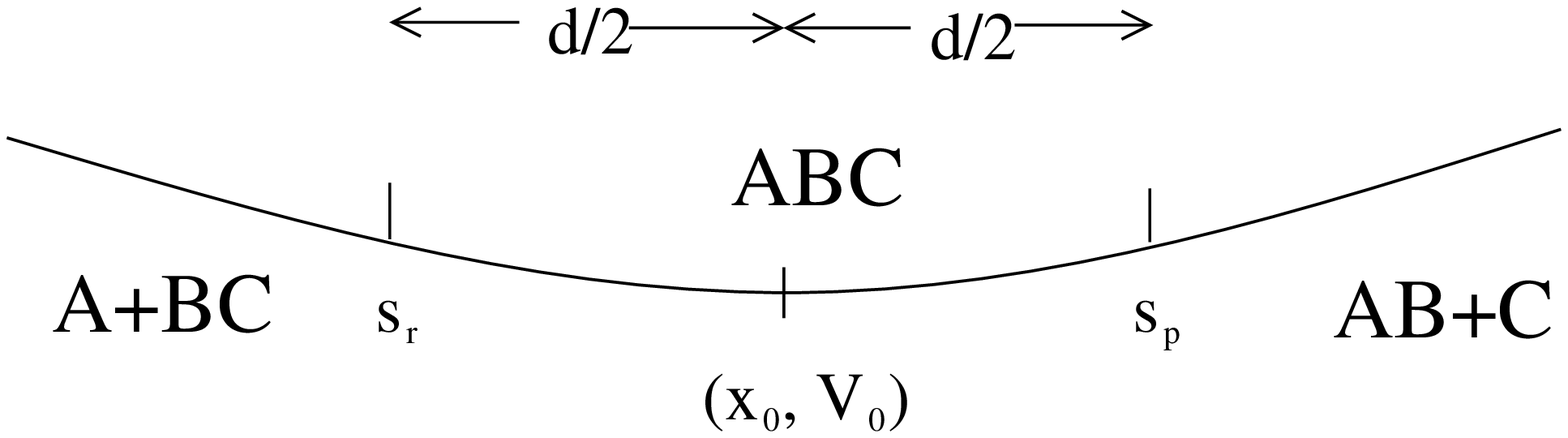,height=12cm}
\caption{K. Saito}
\label{f:hod}
\end{center}
\end{figure}

\begin{figure}[t]
\begin{center}
\epsfig{file=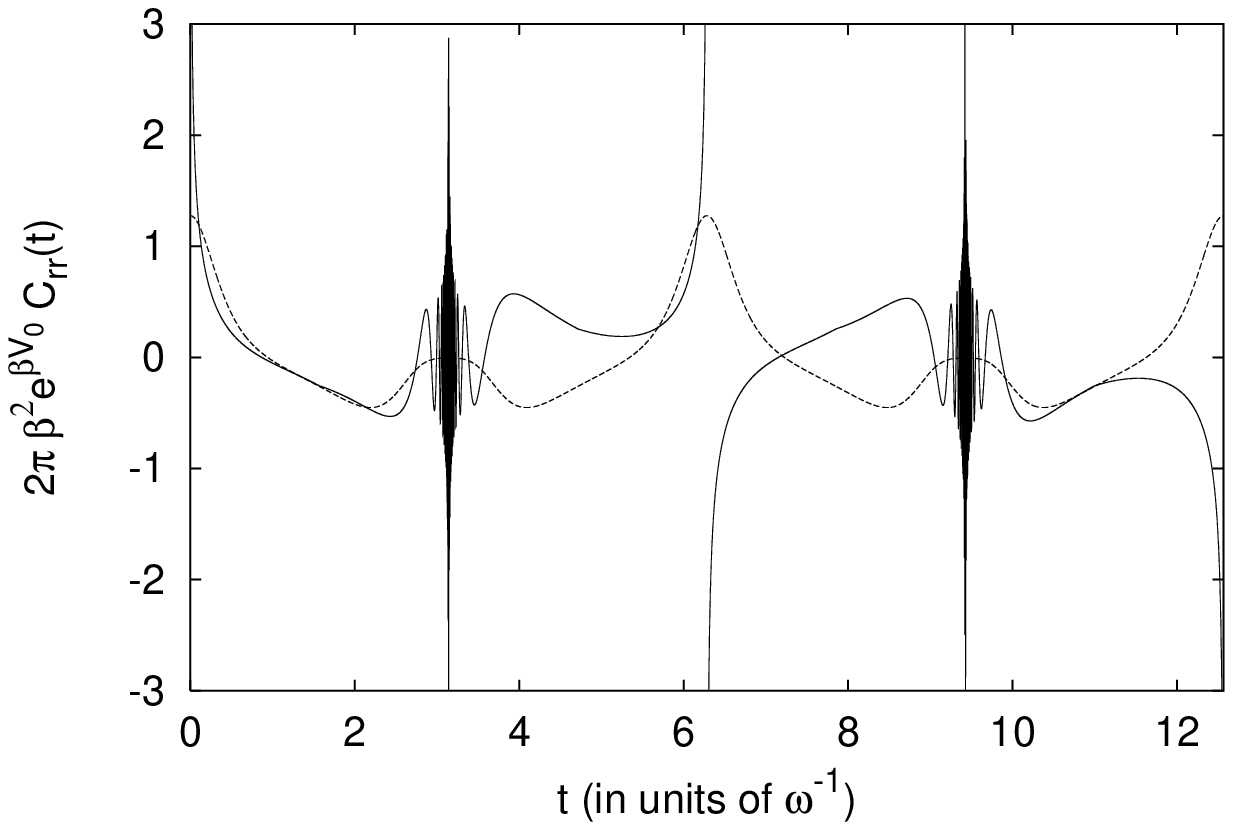,height=11cm}
\caption{K. Saito}
\label{f:horr}
\end{center}
\end{figure}

\begin{figure}[t]
\begin{center}
\epsfig{file=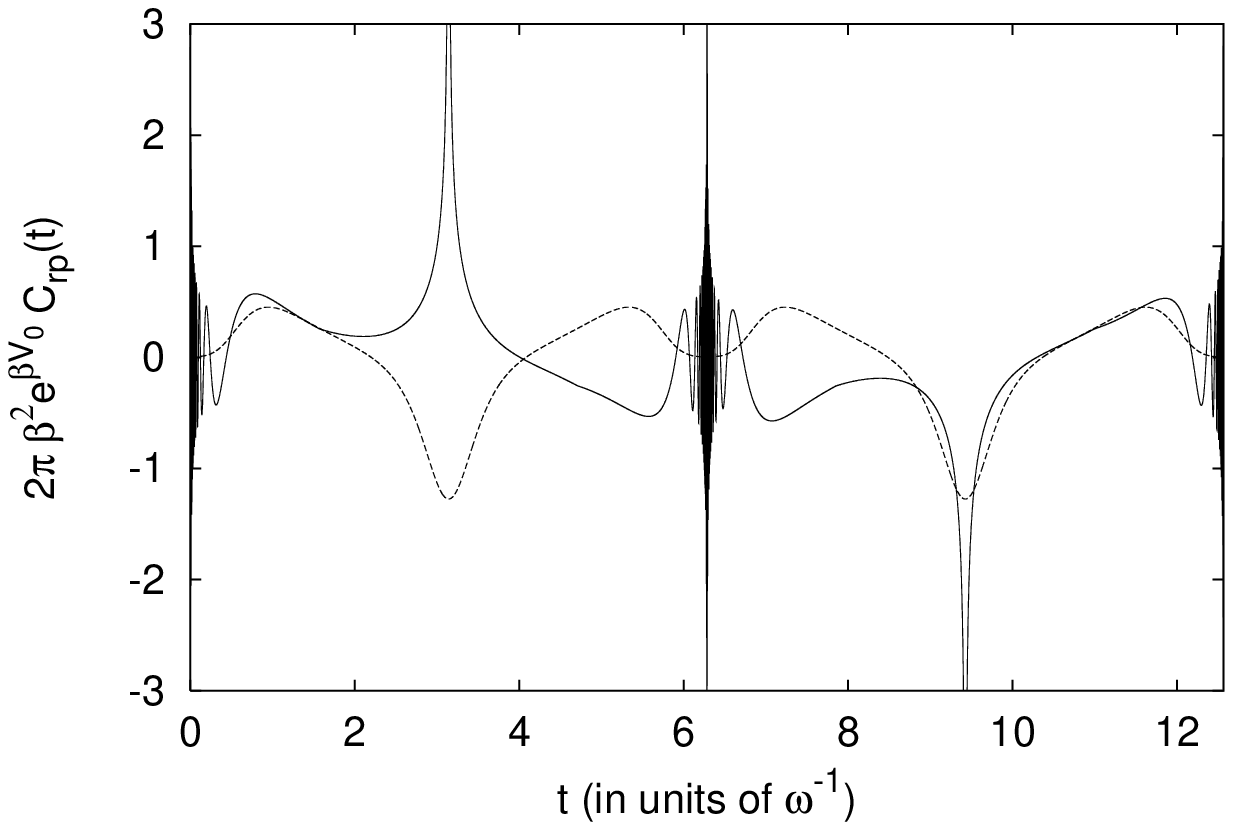,height=11cm}
\caption{K. Saito}
\label{f:horp}
\end{center}
\end{figure}

\noindent Fig. 7: Same as Fig.\ref{f:fRrr} but for the $(rp)$-rate
constant. In the contour plot, the top dotted
curve corresponds to $R_{rp}=0.95$ and the other curves are plotted
at intervals of 0.05. The bottom and right curve is for
$R_{rp}=0.55$. 

\noindent Fig. 8: Same as Fig.\ref{f:fRrr} but for the rate constant
for the recombination reaction. In the contour plot, the top dotted
curve corresponds to $R_{rec}=1.05$ and the other curves are plotted
at intervals of 0.05. The bottom and right curve is for
$R_{rec}=1.35$. 

\noindent Fig. 9: Same as Fig.\ref{f:diag} but for the parabolic
potential case.  The minimum point is located at $(x_0,V_0)$. 

\noindent Fig. 10: Correlation function $C_{rr}(t)$ for the harmonic
oscillator case. The dotted curve is for $C_{rr}^M(t)$, while the
solid curve is for $C_{rr}^Y(t)$. We take $C_1 = 1.5$ and $\kappa =
1.0$. 

\noindent Fig. 11: Correlation function $C_{rp}(t)$ for the harmonic
oscillator case. The dotted curve is for $C_{rp}^M(t)$, while the
solid curve is for $C_{rp}^Y(t)$.  We take $C_1 = 1.5$ and $\kappa =
1.0$. 

\noindent Fig. 12: Ratio of the $(rr)$-rate constants under the
harmonic oscillator potential with $\kappa = 0.05$
(top) and the contour plot (bottom).  In the contour plot, the dotted
curve connecting $(T, P) = (100$K, $0.14$GPa) and $(380$K, $1$GPa)
corresponds to $R_{rr}=1.05$.
The top, left dotted curve connecting $(T, P) = (340$K, $0.1$GPa) and
$(400$K, $0.14$GPa) is also for $R_{rr}=1.05$.
The other curves are plotted
at intervals of 0.05. The bottom and right dotted curve is thus for
$R_{rr}=1.35$. 

\begin{figure}[t]
\begin{center}
\epsfig{file=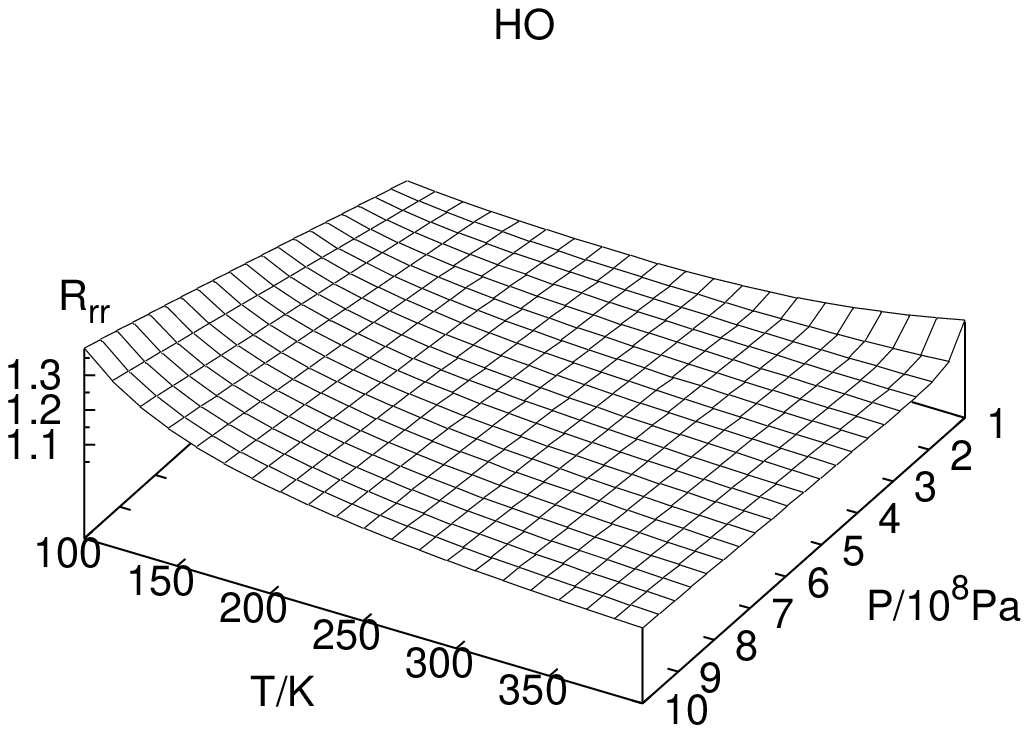,height=10cm}
\epsfig{file=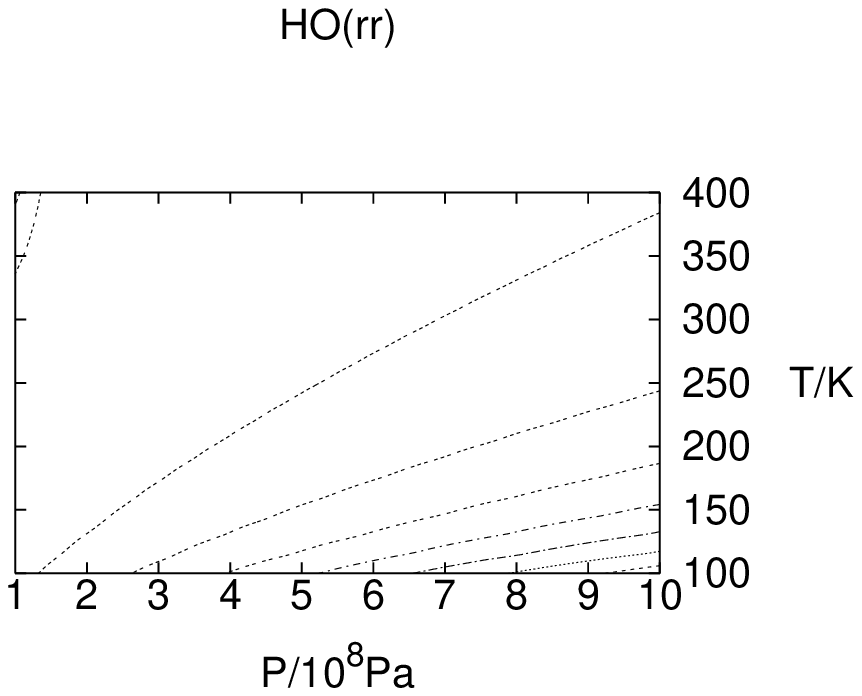,height=10cm}
\caption{K. Saito}
\label{f:hRrr}
\end{center}
\end{figure}

\begin{figure}[t]
\begin{center}
\epsfig{file=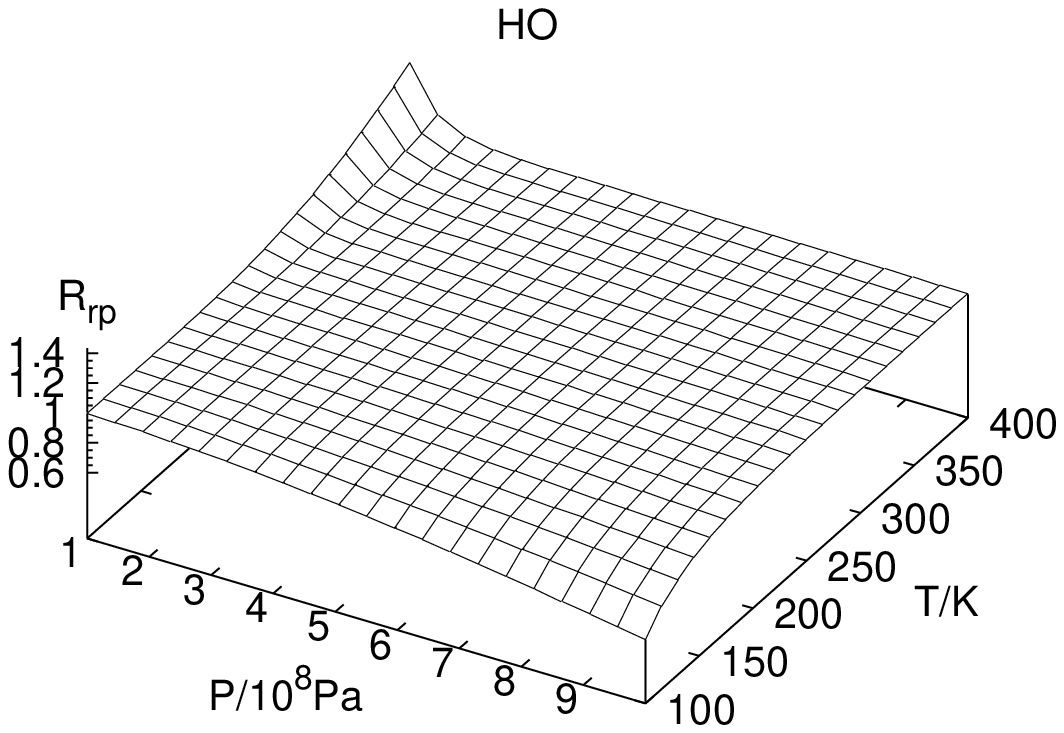,height=10cm}
\epsfig{file=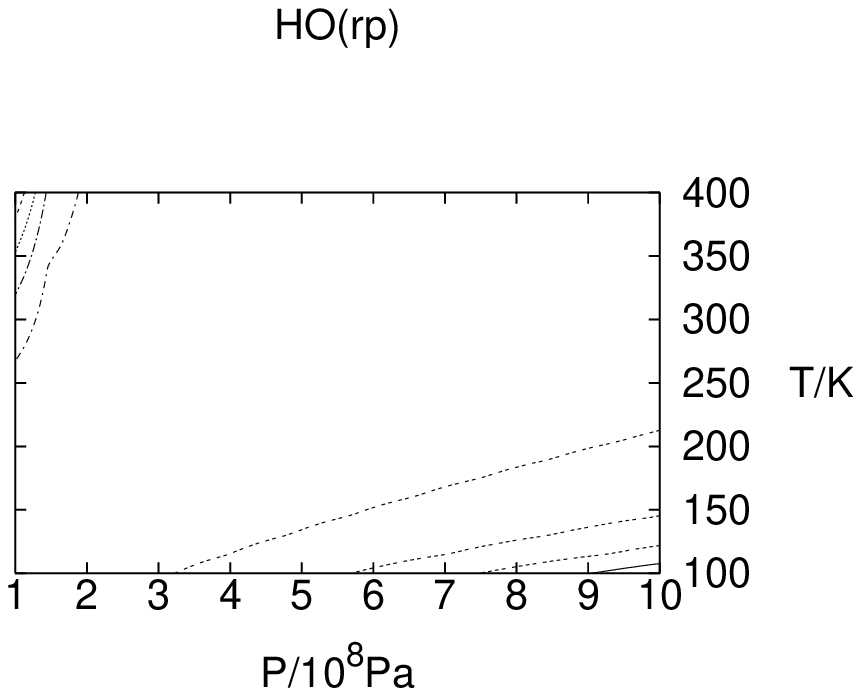,height=10cm}
\caption{K. Saito}
\label{f:hRrp}
\end{center}
\end{figure}

\begin{figure}[t]
\begin{center}
\epsfig{file=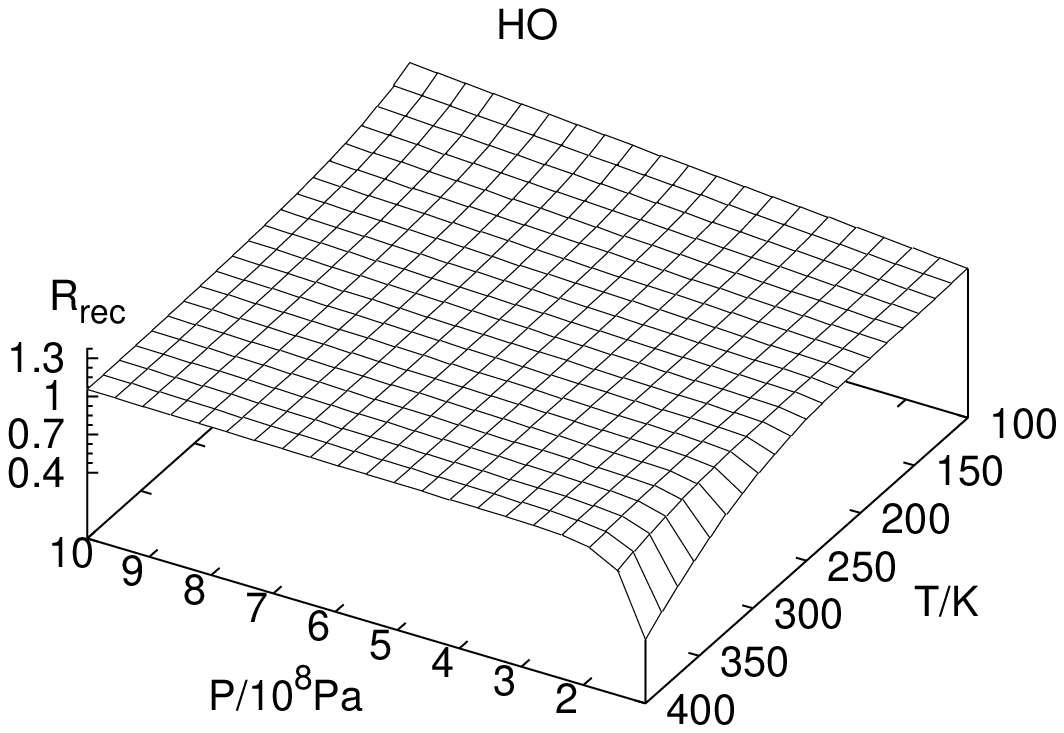,height=10cm}
\epsfig{file=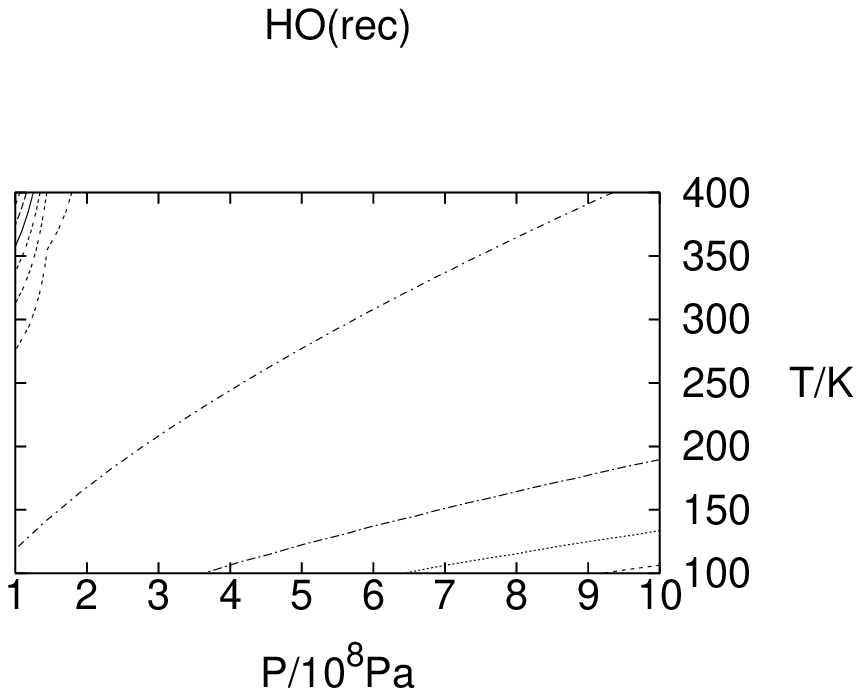,height=10cm}
\caption{K. Saito}
\label{f:hRrec}
\end{center}
\end{figure}

\noindent Fig. 13: Same as Fig.\ref{f:hRrr} but for the $(rp)$-rate
constant. In the contour plot, the dotted curve
connecting $(T, P) = (100$K, $0.33$GPa) and $(210$K, $1$GPa) 
corresponds to $R_{rp}=0.95$, while the dot-dashed curve connecting 
$(T, P) = (270$K, $0.1$GPa) and $(400$K, $0.19$GPa) 
is for $R_{rp}=1.05$.  The other curves are plotted
at intervals of 0.1. The bottom and right curve is for
$R_{rp}=0.65$. 

\noindent Fig. 14: Same as Fig.\ref{f:hRrr} but for the rate constant
for the recombination reaction. In the contour plot, the dot-dashed
curve connecting $(T, P) = (120$K, $0.1$GPa) and $(400$K, $0.93$GPa) 
corresponds to $R_{rec}=1.05$, while the dotted curve connecting 
$(T, P) = (275$K, $0.1$GPa) and $(400$K, $0.18$GPa)
is for $R_{rec}=0.95$.  The other curves are plotted
at intervals of 0.1. The bottom and right curve is for
$R_{rec}=1.35$. 

\newpage

\newpage

\newpage

\newpage

\newpage

\newpage

\newpage

\newpage

\newpage

\newpage

\newpage

\newpage

\newpage

\newpage

\end{document}